%% file: main.tex
\documentclass[floatfix,aps,prd,nofootinbib,superscriptaddress,preprint]{revtex4}

\usepackage{amsmath,amssymb}
\usepackage{bm,bbm}
\usepackage{graphicx, graphics, color}
\usepackage{slashed}
\usepackage{booktabs}
\usepackage{tabu}
\usepackage[dvipsnames]{xcolor}
\usepackage[normalem]{ulem}
\usepackage[colorlinks=true,linkcolor=blue]{hyperref}
\usepackage{tikz}
\usetikzlibrary{shapes,arrows}
\usetikzlibrary{positioning}
\usepackage{cancel}

\newcommand{\fref}[1]{Fig.~\ref{f.#1}}

\newcommand{\eref}[1]{Eq.~(\ref{e.#1})}

\newcommand{\sref}[1]{Section~\ref{s.#1}}

\newcommand{\chis}{\chi^2}

\newcommand{\mell}[1]{{\bm #1}}
\newcommand{\tf}{{\rm T}}

\newcommand{\muR}{\mu_{\rm R}}
\newcommand{\muFF}{\mu_{\rm FF}}
\newcommand{\MD}{\mell{D}}

\preprint{JLAB-THY-16-2327, KEK-TH-1920, J-PARC-TH-0060}

\begin{document}

\title{First Monte Carlo analysis of fragmentation functions \\
	from single-inclusive $e^+ e^-$ annihilation}
\author{N. Sato}
\affiliation{Jefferson Lab, Newport News, Virginia 23606, USA}
\author{J. J. Ethier}
\affiliation{College of William and Mary, Williamsburg, Virginia 23187, USA}
\author{W. Melnitchouk}
\affiliation{Jefferson Lab, Newport News, Virginia 23606, USA}
\author{M. Hirai}
\affiliation{Nippon Institute of Technology, Saitama 345-8501, Japan}
\author{S. Kumano}
\affiliation{High Energy Accelerator Research Organization (KEK),
  1-1, Oho, Tsukuba, Ibaraki 305-0801, Japan}
\affiliation{J-PARC Center, 203-1, Shirakata,
  Tokai, Ibaraki, 319-1106, Japan}
\author{A. Accardi}
\affiliation{Jefferson Lab, Newport News, Virginia 23606, USA}
\affiliation{Hampton University, Hampton, Virginia 23668, USA \\
\vspace*{0.2cm}
{\bf Jefferson Lab Angular Momentum (JAM) Collaboration
\vspace*{0.2cm} }}

\begin{abstract}
We perform the first iterative Monte Carlo (IMC) analysis of
fragmentation functions constrained by all available data from
single-inclusive $e^+ e^-$ annihilation into pions and kaons.
The IMC method eliminates potential bias in traditional analyses
based on single fits introduced by fixing parameters not well
contrained by the data and provides a statistically rigorous
determination of uncertainties.
Our analysis reveals specific features of fragmentation functions
using the new IMC methodology and those obtained from previous
analyses, especially for light quarks and for strange quark
fragmentation to kaons.
\end{abstract}

\date{\today}
\maketitle

\section{Introduction}
\label{s.intro}

Understanding the generation of hadrons from quarks and gluons (partons)
remains a fundamental challenge for strong interaction physics.
High-energy collisions of hadrons or leptons offers the opportunity
to study the formation of mesons and baryons from partons produced in
hard collisions \cite{Albino10, Metz16}.  While the hard scattering
process can be computed perturbatively from the underlying QCD theory,
the hadronization of the quarks and gluons occurs over long distances,
and provides a unique window on nonperturbative QCD
dynamics~\cite{Field78}.

Within the collinear factorization framework \cite{CSS}, the formation
of hadrons is characterized by universal nonperturbative fragmentation
functions (FFs), which in an infinite momentum frame can be interpreted
as probability distributions of specific hadrons $h$ produced with a
fraction $z$ of the scattered parton's longitudinal momentum or energy.
As in the case of parton distribution functions (PDFs), which describe
the quark and gluon momentum distributions inside hadrons, the
nonperturbative FFs are presently not calculable from first principles,
and must be determined phenomenologically from QCD-based analyses of
high-energy scattering data or from QCD-inspired nonperturbative
models \cite{Hrayr11}.

In addition to providing information on the fundamental hadronization
process, FFs are also indispensable tools for extracting information
on the partonic structure of the nucleon from certain high-energy
processes, such as semi-inclusive deep-inelastic scattering (SIDIS)
of leptons from nucleons.  Here, assuming factorization of the
scattering and hadronization subprocesses, the SIDIS cross section
can be expressed in terms of products of PDFs and FFs summed over
individual flavors.
The selection of specific hadrons in the final state, such as
$\pi^\pm$ or $K^\pm$, then allows separation of the momentum and
spin PDFs for different flavors.

The need for well-constrained FFs, especially for kaon production,
has recently been highlighted \cite{LSS10, LSS11, LSS15} in global
analyses of polarized SIDIS observables used to determine the
strange quark contribution $\Delta s$ to the spin of the nucleon.
Inclusive deep-inelastic lepton--nucleon scattering data alone are
incapable of determining this without additional input from theory,
such as the assumption of SU(3) symmetry, or other observables.
Kaon production in polarized SIDIS in principle is such an observable,
involving a new combination of polarized $u$, $d$ and $s$ quark PDFs,
which, when combined with the inclusive data, allow each of the flavor
distributions to be determined -- providing the FFs are known.

As pointed out by Leader {\it et al.} \cite{LSS11}, however, the
variation between the strange-to-kaon FFs from different analyses
is significant and can lead to qualitatively different conclusions
about the magnitude and even sign of the $\Delta s$ distribution.
In particular, analysis \cite{LSS11, DSSV09} of the polarized SIDIS
data using the DSS \cite{DSS07} parametrization of FFs, together
with inclusive DIS polarization asymmetries, suggests a positive
$\Delta s$ at intermediate $x$ values, $x \sim 0.1-0.2$, in contrast
to the generally negative $\Delta s$ at all $x$ obtained from
inclusive DIS data alone, assuming constraints on the weak baryon
decays from SU(3) symmetry \cite{JAM15}.
Employing instead the HKNS \cite{HKNS07} FF parametrization, in which
the strange fragmentation to kaons is several times smaller in some
regions of $z$ compared with that from the DSS \cite{DSS07} fit,
yields a negative $\Delta s$ consistent with the inclusive-only
analyses \cite{LSS15}.
It is crucial, therefore, to understand the origin of the differences
in the magnitudes and shapes of the strange, as well as other,
FFs found in the different analyses before one can draw reliable
conclusions about the strange quark content of the nucleon extracted
from analyses including SIDIS data.

Differences between FFs can come from a variety of sources, including
different data sets used in the analyses (single-inclusive $e^+ e^-$
annihilation, SIDIS, inclusive hadron production in $pp$ collisions),
the choice of parametrization for the FFs, assumptions about FFs
that are not well constrained by data, or even the presence of local
minima in the fitting procedure.
Most of the analyses to date have been performed at next-to-leading
order (NLO) accuracy in the strong coupling constant \cite{Kretzer00,
AKK05, AKK06, AKK08, HKNS07, DSS07, DSS15, LSS10, LSS11, LSS15,
Hirai16}, although more recent studies have explored the effects
of incorporating next-to-next-to-leading order (NNLO) corrections
\cite{ASR15}, as well as other theoretical developments such as
threshold resummation \cite{AKKresum, ARV13, AAR15} and hadron mass
effects~\cite{AAR15}.

A common feature of all existing FF analyses is that they are
obtained from single fits, using either $e^+ e^-$ single-inclusive
annihilation (SIA) data alone, or in combination with unpolarized
SIDIS and inclusive hadron production in $pp$ collisions.
In order to address some of the questions raised by the recent
ambiguities in the strange quark FFs and their impact on the
$\Delta s$ determination, in this paper we go beyond the standard
fitting paradigm by performing the first Monte Carlo (MC) analysis
of FFs.  In particular, we extend the methodology of the iterative
Monte Carlo (IMC) approach introduced in Ref.~\cite{JAM15} for the
analysis of spin-dependent PDFs to the case of FFs.

The virtue of the IMC approach is that it allows for a full
exploration of the parameter space when sampling initial priors
for any chosen parametric form for the fitting function.
It thereby eliminates any bias introduced by fine-tuning or fixing
specific parameters that are not well contrained by the data,
a practice often employed to control single fits.
Furthermore, the conventional polynomial-type parametrization
choice can have multiple solutions that lead to various local
minima in the $\chi^2$ landscape, whereas the IMC technique
statistically surveys all possible solutions, thereby avoiding
the fit being stuck in false minima.

A further important advantage of the IMC technology is in the
extraction of uncertainties on the FFs.  In standard analyses the
theoretical errors are typically determined using the Hessian
\cite{HKNS07} or Lagrange multiplier methods \cite{DSS07},
in which a tolerance parameter $\Delta\chi^2$ is introduced
to satisfy a specific confidence level (CL) of a $\chi^2$
probability density function with $N$ degrees of freedom.
In the IMC framework, the need for tolerance criteria is
eliminated entirely and the uncertainties are extracted through
a robust statistical analysis of the Monte Carlo results.

As a first IMC analysis of FFs, we confine ourselves to the
case of charged pion and kaon production in $e^+ e^-$ SIA, using
all available $\pi^\pm$ and $K^\pm$ cross section data from
DESY \cite{TASSO80, TASSO83, TASSO89, ARGUS89},
SLAC \cite{TPC84, TPC86, TPC88, HRS87, SLD04},
CERN \cite{OPAL94, OPAL00, ALEPH95, DELPHI95, DELPHI98}, and
KEK \cite{TOPAZ95}, as well as more recent, high-precision
results from the Belle \cite{Belle13, Leitgab13} and BaBar
\cite{BaBar13} Collaborations at KEK and SLAC, respectively.
Although SIA data in principle only constrain the sum of the quark
and antiquark distributions, we also make use of flavor-tagged data
\cite{OPAL00} which allow separation of hadron production from heavy
and light quarks.
In addition, the availability of data over a range of kinematics,
from relatively low center-of-mass energies $Q \approx 10$~GeV up to
the $Z$-boson pole, $Q \approx 91$~GeV, allows for the separation
of the up- and down-type FFs due to differences in the quark--boson
couplings in the $\gamma$ and $Z$ channels \cite{Hirai16}.
To ensure proper treatment of data at $z \sim 1$, we systematically
apply correct binning by integrating over each $z$ bins, rather than
taking bin averages as in previous analyses.  We also studied the
$z$ cuts on the data in different channels that need to be applied
at low $z$ values, below which the collinear framework breaks
down and our analysis is not expected to be reliable.

Note that our aim here is not so much the definitive determination
of FFs, which would require inclusion of all possible processes
that have sensitivity to FFs, but rather to explore the application
of the IMC methodology for FFs to determine the maximal information
that can be extracted from the basic $e^+ e^-$ SIA process alone.
The lessons learned here will be used in subsequent analyses of the
entire global set of SIA and other high-energy scattering data to
provide a more definitive determination of the individual FFs.

We begin in Sec.~\ref{s.theory} by reviewing the formalism for the
$e^+ e^-$ annihilation into hadrons, including a summary of the
SIA cross sections at NLO and $Q^2$ evolution of the fragmentation
functions.  To improve the computational efficiency we perform the
numerical calculations in moment space, recontructing the momentum
dependence of the fragmentation functions using inverse Mellin
transforms.
The methodology underpinning our global analysis is presented in
Sec.~\ref{s.method}, where we describe the parametrizations
employed and the treatment of uncertainties.
This section also outlines the essential features of the IMC
method used to perform the fits to the data, highlighting several
improvements in the methodology compared to that introduced 
originally in the global analysis of the JAM spin-dependent
PDFs~\cite{JAM15}.
The experimental data sets analyzed in this study are summarized
in Sec.~\ref{s.data}, and the results of our analysis presented
in Sec.~\ref{s.results}.  We compare the fitted cross sections
with all available $e^+ e^-$ data, for both inclusive and
flavor-tagged cross sections, finding good overall $\chi^2$
values for both pion and kaon production.
We illustrate the convergence of the iterative procedure for
the favored and unfavored FFs, the latter being partially
constrained by the flavor-tagged data.
The shapes and magnitudes of the FFs from our IMC analysis are
compared and contrasted with those from previous global fits,
highlighting important differences in the light quark sector
and for quark fragmentation to kaons.
Finally, in Sec.~\ref{s.conclusion} we summarize our findings
and preview future extensions of the present analysis.

\section{Formalism}
\label{s.theory}

\subsection{Cross section and fragmentation functions}
\label{ss.crosssection}

The $e^+ e^- \to h X$ cross section is typically measured as a
function of the variable \mbox{$z=2p_h\cdot q/Q^2$}, where $p_h$ is
the momentum of the detected hadron $h$ and $q$ is the momentum of
the exchanged photon or $Z$-boson with invariant mass $Q=\sqrt{Q^2}$.
In the $e^+ e^-$ center-of-mass frame, $z=2E_h/Q$ can be interpreted
as the momentum fraction of the parent quark carried by the produced
hadron.  For a given hadron $h$ the experimental $z$ distribution is
usually given as
\begin{align}
F^h(z,Q^2) = \frac{1}{\sigma_{\rm tot}} \frac{d\sigma^h}{dz}(z,Q^2),
\label{e.empirical}
\end{align}
which we shall refer to as the \emph{empirical} fragmentation function
for a given hadron of type $h$.  In Eq.~(\ref{e.empirical}) the total
inclusive $e^+ e^-$ cross section $\sigma_{\rm tot}$ can be calculated
at NLO as
\begin{align}
\sigma_{\rm tot}(Q^2)
= \sum_q \frac{4\pi \alpha^2}{Q^2} \tilde{e}_q^2
  \left(1 + 4 a_s(\muR^2)\right)
+ {\cal O}(a_s^2),
\label{e.sigtot}
\end{align}
where $\alpha = e^2/4\pi$ is the electromagnetic fine structure
constant and $a_s(\muR) \equiv \alpha_s(\muR)/4\pi$, with the strong
coupling constant $\alpha_s$ evaluated at the ultraviolet
renormalization scale $\muR$.
The index $q$ runs over the active quark flavors allowed by the hard
scale $Q$, and we introduce the shorthand notation for the charges
\begin{align}
\tilde{e}_q=
   e_q^2 + 2 e_q\, g_V^q g_V^e\, \rho_1(Q^2)
   + \left( g_A^{e\, 2} + g_V^{e\, 2} \right)
     \left( g_A^{q\, 2} + g_V^{q\, 2} \right)
     \rho_2(Q^2).
\end{align}
Here the quark vector and axial vector couplings are given by
  $g_V^q = \frac{1}{2} - \frac{4}{3}\sin^2\theta_W$
and
  $g_A^q = +\frac{1}{2}$
for the $q=u,c$ flavors, while for the $q=d,s,b$ flavors
these are
  $g_V^q = -\frac{1}{2} + \frac{2}{3}\sin^2\theta_W$
and
  $g_A^q = -\frac{1}{2}$.
Similarly, the electron vector and axial vector couplings are
given by
  $g_V^e = -\frac{1}{2} + 2\sin^2\theta_W$
and
  $g_A^e = -\frac{1}{2}$,
respectively.
Because the weak mixing angle $\sin^2\theta_W$ is $\approx 1/4$,
the contribution from the vector electron coupling is strongly
suppressed relative to the axial vector coupling.
The terms with $\rho_1$ and $\rho_2$ arise from $\gamma Z$
interference and $Z$ processes, respectively, and are given by
\begin{subequations}
\begin{eqnarray}
\rho_1(Q^2)
&=& \frac{1}{4 \sin^2\theta_W \cos^2\theta_W}
    \frac{Q^2 (M_Z^2-Q^2)}{(M_Z^2-Q^2)^2 + M_Z^2 \Gamma_Z^2},	\\
\rho_2(Q^2)
&=& \frac{1}{\left( 4 \sin^2\theta_W \cos^2\theta_W \right)^2}
    \frac{Q^4}{(M_Z^2-Q^2)^2 + M_Z^2 \Gamma_Z^2},
\end{eqnarray}
\end{subequations}
where $M_Z$ and $\Gamma_Z$ are the mass and width of the $Z$ boson,
respectively.

Within the collinear factorization framework, the empirical
fragmentation function $F^h(z,Q^2)$ can be approximately calculated
in terms of quark fragmentation functions into hadrons,
\begin{align}
F^h(z,Q^2) \approx F_{\rm coll}^h(z,Q^2) =
\sum_i \left[H_i \otimes D^h_i \right](z,Q^2,\muR^2,\muFF^2)
 + {\cal O}(a_s^2),
\label{e.F}
\end{align}
where ``$\otimes$'' refers to the standard convolution integral
$[H \otimes D](z)
 = \int_z^1 (d\hat{z}/\hat{z})\, H(\hat{z}) D(z/\hat{z})$,
and the sum runs over all parton flavors $i = q, \bar{q}, g$.
Here $H_i$ is the short-distance hard cross section calculable
in fixed-order perturbative QCD, and $D_i^h$ is the partonic
fragmentation function.  As discussed below, the quark contributions
$H_q$ depend on the charges $\tilde{e}^2_q$, while the gluon
contribution is independent of the charges.

At NLO in the $\overline{\rm MS}$ scheme (which we use throughout
in this analysis), the hard cross section can be written
\begin{align}
H_i(\hat{z},Q^2,\muR^2,\muFF^2)
= H_i^{(0)}(\hat{z},Q^2,\muR^2,\muFF^2) 
+ a_s(\muR)\,
  H_i^{(1)}(\hat{z},Q^2,\muR^2,\muFF^2) 
+ \mathcal{O}(a_s^2),
\label{e.hard}
\end{align}
where $\hat{z}$ is the partonic energy fraction carried
by the outgoing hadron.
As in Eq.~(\ref{e.sigtot}), $\muR$ is the
renormalization scale stemming from regularization of the
ultraviolet divergences in the virtual graphs that contribute to
$H_i^{(1)}$, while $\muFF$ is a factorization scale associated
with the FF $D_i^h$.
Note that the dependence of the convolution integral in
Eq.~(\ref{e.F}) on the scales $\muR$ and $\muFF$ is a remnant of
the fixed-order perturbative QCD approximation to $F_{\rm coll}$,
which will be cancelled by inclusion of higher order terms in the
perturbative series.
At leading order in $a_s$, the $2 \to 2$ phase space is
such that $\hat{z}=z$, so that $H_i^{(0)}$ is proportional to
$\delta(\hat{z}-z)$.
At higher orders, additional QCD radiation effects open up
the phase space for the outgoing fragmenting parton such that
$\hat{z}$ varies between $z$ and 1.

The partonic FF $D_i^h$ can be interpreted as the number density to
find a hadron of type $h$ in the jet originating from the parton $i$
with momentum fraction $\hat{z}$ \cite{CollinsBook}.  As for PDFs,
FFs are sensitive to ultraviolet divergences, and after renormalization
they acquire dependence on the scale $\muFF$.  (The subscript
``FF'' denotes the {\it final} state factorization scale, in
contrast to the {\it initial} state factorization scale in PDFs.)
In practice, to optimize the perturbative expansion of the hard
cross section, we set $\muR=\muFF=Q$.  However, for completeness
we leave the dependence of $\muR$ and $\muFF$ in \eref{F} and below
explicit.  In general, variation of the scales around $Q$ allows
one to assess the uncertainty in the perturbative expansion.
For instance, in Ref.~\cite{ASR15} a significant reduction of
the scale dependence was found with the inclusion of the NNLO
corrections.

\subsection{Scale dependence}
\label{ss.evolution}

In perturbative QCD the scale dependence of the FFs is described
by the evolution equations,
\begin{align}
\frac{d D^h_i(\hat{z},\muFF^2)}{d\ln(\muFF^2)}
= \left[ P_{ij} \otimes D_j^h \right](\hat{z},\muFF^2),
\label{e.dglap}
\end{align}
where $P_{ij}$ are the timelike $i \to j$ splitting functions.
Since the FFs cannot be calculated from first principles, the
$\hat{z}$ dependence is fitted to the data at some input scale
$\muFF^2=Q^2_0$.  The latter is chosen at the lowest possible value
where a perturbative QCD description can be applied in order to
minimize errors induced by backward evolution from the truncation
of the perturbative series.

The simplest approach to solving the evolution equations
(\ref{e.dglap}) is to use one of several numerical approximation
techniques to solve the integro-differential equations directly in
$\hat{z}$ space \cite{Hirai12}.  Alternatively, as discussed in
Ref.~\cite{JAM15}, it can be more efficient to solve the equations
in Mellin moment space, where the $N$-th Mellin moment of a function
$f(z)$ is defined as
\begin{eqnarray}
\mell{f}(N)
&=& \int_0^1 dz\, z^{N-1}\, f(z),
\end{eqnarray}
and similarly for all other moments of functions denoted in boldface.
In this framework the convolution integrals in Eqs.~(\ref{e.hard})
and (\ref{e.dglap}) can be rendered as ordinary products of the
Mellin moments,
\begin{align}
\mell{F}^h_{\rm coll}(N,Q^2)
&= \sum_i \mell{H_i}(N,Q^2,\muR^2,\muFF^2)\, 
	  \mell{D^h_i}(N,Q^2,\muR^2,\muFF^2)
 + {\cal O}(a_s^2),
\label{e.Fmoments}
\end{align}
and
\begin{align}
\frac{d\mell{D}^h_i(N,\muFF^2)}{d\ln(\muFF^2)}
&= \mell{P}_{ij}(N,\muR^2,\muFF^2)\,
   \mell{D}_{j}^h(N,Q^2,\muR^2,\muFF^2).
\label{e.mom_evol}
\end{align}
The evolution equations for $\mell{D}_i^h$ can be solved using
the methods described in Ref.~\cite{PEGASUS}, and the hadronic
fragmentation function in $z$-space can be obtained using the
inverse Mellin transform,
\begin{align}
F^h_{\rm coll}(z,Q^2)
&= \frac{1}{2\pi i} \int_C dN\, z^{-N}\, \mell{F}^h_{\rm coll}(N,Q^2).
\label{e.invmell}
\end{align}
The main advantage of the Mellin techniques is the improvement in
speed in the evaluation of the observables and evolution equations.
Another advantage is that the experimental cross sections are
typically presented as averaged values over bins of $z$.
Such averaging, between $z_{\rm min}$ and $z_{\rm max}$, can be
simply done analytically,
\begin{align}
\left<F^h_{\rm coll}(z,Q^2)\right>_{z\, {\rm bin}}
= \frac{1}{\left( z_{\rm max}-z_{\rm min} \right)}
  \frac{1}{2\pi i}\int_C dN\,
  \frac{\left( z^{1-N}_{\rm max} - z^{1-N}_{\rm min} \right)}{1-N}
  \mell{F}^h_{\rm coll}(N,Q^2),
\label{e.invmellavg}
\end{align}
without deteriorating the numerical performance.
In contrast, such advantage does not exist if one evaluates
$F^h_{\rm coll}(z,Q^2)$ and solves the DGLAP evolution equations
directly in $z$ space \cite{SV01}.  In practice, at small $z$ the
bins sizes are quite small and taking the central $z$ values
might be appropriate.  However, at large $z$ the bin sizes
increase and, depending on the precision of the measured cross 
sections, the averaging step becomes important.

For clarity, we express the Mellin moments of the hard factor
in \eref{Fmoments} in terms of unnormalized hard factors
$\widetilde{\mell{H}}_i$,
\begin{subequations}
\begin{align}
\mell{H}_q(N,Q^2,\muR^2,\muFF^2)
&= \frac{\tilde{e}_q^2}{\sum_{q'} \tilde{e}_{q'}^2}\
   \frac{\widetilde{\mell{H}}_q(N,Q^2,\muR^2,\muFF^2)}
	{\left( 1 + 4 a_s(\muR^2) \right)},		\\
\mell{H}_g(N,Q^2,\muR^2,\muFF^2)
&= \frac{\widetilde{\mell{H}}_g(N,Q^2,\muR^2,\muFF^2)}
	{\left( 1 + 4 a_s(\muR^2) \right)},
\end{align}
\end{subequations}
where the charge factors for the gluon moments cancel.
The perturbative expansion of $\widetilde{\mell{H}}_i$ is
then given by
\begin{subequations}
\begin{align}
\widetilde{\mell{H}}_q(N,Q^2,\muR^2,\muFF^2)
& = 1 + a_s(\muR^2)\,
	\widetilde{\mell{H}}^{(1)}_q(N,Q^2,\muR^2,\muFF^2) 
  + \mathcal{O}(a_s^2),			\\
\widetilde{\mell{H}}_g(N,Q^2,\muR^2,\muFF^2)
& = a_s(\muR^2)\,
    \widetilde{\mell{H}}^{(1)}_g(N,Q^2,\muR^2,\muFF^2)
  + \mathcal{O}(a_s^2),
\end{align}
\end{subequations}
where the gluon contribution begins at NLO.  Physically, this
corresponds to gluon fragmentation into hadrons from real QCD
radiation that occurs at NLO.  For completeness, in Appendix~A
we list the formulas for $\widetilde{\mell{H}}^{(1)}_{q,g}$
at NLO.

To solve the evolution equations in \eref{Fmoments}, we follow
the conventions of Ref.~\cite{PEGASUS}, which we briefly summarize
here.  For convenience we work in a flavor singlet and nonsinglet
basis, in which we define the flavor combinations
\begin{subequations}
\begin{align} 
\mell{D}^{h}_{\pm3} 
  &=\MD^h_{u^{\pm}}
   -\MD^h_{d^{\pm}},	\\
\mell{D}^{h}_{\pm8}
  &=\MD^h_{u^{\pm}}
   +\MD^h_{d^{\pm}}
   -2\MD^h_{s^{\pm}},	\\
\mell{D}^{h}_{\pm15}
  &=\MD^h_{u^{\pm}}
   +\MD^h_{d^{\pm}}
   +\MD^h_{s^{\pm}}
   -3\MD^h_{c^{\pm}},	\\
\mell{D}^{h}_{\pm24}
  &=\MD^h_{u^{\pm}}
   +\MD^h_{d^{\pm}}
   +\MD^h_{s^{\pm}}
   +\MD^h_{c^{\pm}}
   -4\MD^h_{b^{\pm}},	\\
\mell{D}^{h}_{\pm35}
  &=\MD^h_{u^{\pm}}
   +\MD^h_{d^{\pm}}
   +\MD^h_{s^{\pm}}
   +\MD^h_{c^{\pm}}
   +\MD^h_{b^{\pm}}
   -5\MD^h_{t^{\pm}},	\\
\mell{D^h_{\pm}} 
  &= \MD^h_{u^{\pm}}
   +\MD^h_{d^{\pm}}
   +\MD^h_{s^{\pm}}
   +\MD^h_{c^{\pm}}
   +\MD^h_{b^{\pm}}
   +\MD^h_{t^{\pm}} \ ,
\end{align}
\end{subequations}
where $\MD_{q^\pm}^h$ are the Mellin moments of the charge
conjugation-even and -odd FFs
$D^h_{q^{\pm}}(z,Q^2) = D^h_q(z,Q^2) \pm D^h_{\bar q}(z,Q^2)$.
Depending on the number of active flavors $n_f$, one needs to
consider only the equations up to $\mell{D}^{\pm}_{n_f^2-1}$,
otherwise the system becomes degenerate.  The evolution equations
in this basis can be expressed as
\begin{subequations}
\begin{align}
\frac{\partial \mell{D}^{h}_{\pm j}}{\partial\ln \muFF^2}
&= \mell{P}_{\rm NS}^{\pm}\,
   \mell{D}^{h}_{\pm j},				\\
\frac{\partial \mell{D}^h_{-}}{\partial\ln \muFF^2}
&= \mell{P}_{\rm NS}^{-}\,
   \mell{D}^{h}_-					\\
\frac{\partial}{\partial\ln\muFF^2}
   \begin{pmatrix}
   \mell{D}^h_{+} \\ \mell{D}^h_{g}
   \end{pmatrix}
&= \begin{pmatrix}
   \mell{P}_{qq}&\mell{P}_{qg} \\ \mell{P}_{gq}&\mell{P}_{gg}
   \end{pmatrix}
   \begin{pmatrix}
   \mell{D}^h_{+} \\ \mell{D}^h_{g}
   \end{pmatrix},
\end{align}
\end{subequations}
with the splitting functions in Mellin space $\mell{P}_{ij}$
listed in Appendix~B.
An important observation here is that all the ``$+$'' FFs maximally
couple to the gluon FFs, while the ``$-$'' functions decouple
completely.  In particular, if one consider observables that depend
only on ``$+$'' combinations, then the ``$-$'' components can be
ignored.  

In our analysis we use an independent implementation
of the evolution equations in Mellin space as described in
Ref.~\cite{PEGASUS}, finding excellent agreement with existing
evolution codes.

\section{Methodology}
\label{s.method}


\subsection{Input scale parametrization }
\label{ss.parametrization}

In choosing a functional form for the FFs, it is important to
note that the SIA observables are sensitive only to the charge
conjugation-even quark distributions $D^h_{q^+}(z,Q^2)$
and the gluon FF $D^h_g(z,Q^2)$.
These couple maximally in the $Q^2$ evolution equations, while
the charge conjugation-odd combinations $D^h_{q^-}(z,Q^2)$
decouple entirely from both $D^h_{q^+}(z,Q^2)$ and $D^h_g(z,Q^2)$.
In our analysis we therefore seek only to extract the $D^h_{q^+}$
and gluon distributions, and do not attempt to separate quark and
antiquark FFs.  This would require additional data, such as from
semi-inclusive deep-inelastic hadron production, which can provide
a filter on the quark and antiquark flavors.

As a reference point, we consider a ``template'' function of the form
\begin{align}
\tf(z;\bm{a})
= M \frac{z^\alpha (1-z)^\beta}
	 {\int_0^1 dz\, z^{1+\alpha}(1-z)^\beta},
\label{e.Tza}
\end{align}
where $\bm{a}=\{M,\alpha,\beta\}$ is the vector of shape parameters
to be fitted.  The denominator is chosen so that the coefficient
$M$ corresponds to the average momentum fraction $z$.

Using charge conjugation symmetry, one can relate
\begin{align}
D^{h^+}_{q^+} = D^{h^-}_{q^+}, \qquad
D^{h^+}_g = D^{h^-}_g,
\end{align}
for all partons.  For pions we further use isospin symmetry to set
the $u^+$ and $d^+$ functions equal, while keeping the remaining FFs
independent.  Since the $u^+$ and $d^+$ distributions must reflect
both the ``valence'' and ``sea'' content of the $\pi^+$, we allow
two independent shapes for these, while a single template function
should be sufficient for the heavier flavors and the gluon,
\begin{subequations}
\label{e.piparam}
\begin{align}
D^{\pi^+}_{u^+}
 = D^{\pi^+}_{d^+}
&= \tf(z;\bm{a}_{ud}^\pi) + \tf(z;\bm{a}_{ud}^{\prime\pi}),	\\
D^{\pi^+}_{s^+,\, c^+,\, b^+,\, g}
&= \tf(z;\bm{a}_{s,\, c,\, b,\, g}^\pi).
\end{align}
\end{subequations}
The additional template shape for the $u^+$ or $d^+$ increases
the flexibility of the parametrization in order to accomodate the
distinction between favored (``valence'') and unfavored (``sea'')
distributions, having different sets of shape parameters
$\bm{a}_{ud}^\pi$ and $\bm{a}_{ud}^{\prime \pi}$.

For the kaon the $s^+$ and $u^+$ FFs are parametrized independently
because of the mass difference between the strange and up quarks.
Since these contain both valence and sea structures, to improve the
flexibility of the parametrization we use two template shapes here,
and one shape for each of the other distributions,
\begin{subequations}
\label{e.Kparam}
\begin{align}
D^{K^+}_{s^+}
&= \tf(z;\bm{a}_s^K) + \tf(z;\bm{a}_s^{\prime K}),	\\
D^{K^+}_{u^+}
&= \tf(z;\bm{a}_u^K) + \tf(z;\bm{a}_u^{\prime K}),	\\
D^{K^+}_{d^+,\, c^+,\, b^+,\, g}
&= \tf(z;\bm{a}_{d,\, c,\, b,\, g}^K).
\end{align}
\end{subequations}
The total number of free parameters for the kaon FFs is 24,
while for the pions the number of parameters is 18.

For the heavy quarks $c$ and $b$ we use the zero-mass variable
flavor scheme and activate the heavy quark distributions at their
mass thresholds, $m_c=1.43$~GeV and $m_b=4.3$~GeV.
For the $Q^2$ evolution we use the ``truncated'' solution in
Ref.~\cite{PEGASUS}, which is more consistent with fixed-order
calculations.  Finally, the strong coupling is evaluated by
solving numerically the $\beta$-function at two loops and using
the boundary condition at the $Z$ pole,
$\alpha_s(m_Z) = 0.118$.

\subsection{Iterative Monte Carlo fitting}
\label{s.imc}

In all previous global analyses of FFs, only single $\chi^2$ fits
have been performed.  In this case it is common to fix by hand
certain shape parameters that are difficult to constrain by data
in order to obtain a reasonable fit.  However, since some of the
parameters and distributions are strongly correlated, this can bias
the results of the analysis.  In addition, there is no way to
determine {\it a priori} whether a single $\chi^2$ fit will become
stuck in any one of many local minima.
The issues of multiple solutions can be efficiently avoided through
MC sampling of the parameter space, which allows exploration of all
possible solutions.
Since this study is the first MC-based analysis of FFs, we briefly
review the IMC procedure, previously introduced in the JAM15
analysis of polarized PDFs \cite{JAM15}, and highlight several
important new features.

In the IMC methodology, for a given observable $\mathcal{O}$ the
expectation value and variance are defined by
\begin{align}
{\rm E}[\mathcal{O}] &= \int d^m a\, \mathcal{P}(\bm{a}|{\rm data})\,
\mathcal{O}(\bm{a}),
\label{e.E}					\\
{\rm V}[\mathcal{O}] &= \int d^m a\, \mathcal{P}(\bm{a}|{\rm data})\,
(\mathcal{O}\left( \bm{a})-{\rm E}[\mathcal{O}] \right)^2,
\label{e.V}
\end{align}
respectively, where $\bm{a}$ is the $m$-component vector representing
the shape parameters of the FFs.  The multivariate probability
density $\mathcal{P}(\bm{a}|{\rm data})$ for the parameters $\bm{a}$
conditioned by the evidence ({\it e.g.}, the data) can be written as 
\begin{align}
\mathcal{P}(\bm{a}|{\rm data})\
\propto\
\mathcal{L}({\rm data}|\bm{a})\times\pi(\bm{a}),
\label{e.P}
\end{align}
where $\pi(\bm{a})$ is the \emph{prior} and
$\mathcal{L}({\rm data}|\bm{a})$ is the \emph{likelihood}.
In our analysis $\pi(\bm{a})$ is initially set to be a flat
distribution.  For $\mathcal{L}({\rm data}|\bm{a})$ we assume
a Gaussian likelihood,
\begin{align}
\mathcal{L}({\rm data}|\bm{a})\
\propto\ \exp\left(-\frac{1}{2} \chi^2(\bm{a})\right),
\end{align}
with the $\chi^2$ function defined as
\begin{align}
\chi^2(\bm{a}) &= \sum_e
\left[ \sum_i
  \left( \frac{{\cal D}_i^{(e)} N^{(e)}_i - T^{(e)}_i}
	      {\alpha^{(e)}_iN^{(e)}_i}
  \right)^2
+ \sum_k \left(r^{(e)}_k\right)^2
\right].
\label{e.chi2}
\end{align}
Here ${\cal D}_i^{(e)}$ and $T^{(e)}_i$ represent the data and theory
points, respectively, and $\alpha^{(e)}_i$ are the uncorrelated
systematic and statistical experimental uncertainties added in
quadrature.  The normalization uncertainties are accounted for
through the factor $N^{(e)}_i$, defined as
\begin{align}
N_i^{(e)}
= 1 - \sum_k\frac{r_k^{(e)} \beta_{k,i}^{(e)}}{{\cal D}_i^{(e)}}.
\end{align}
Here $\beta_{k,i}^{(e)}$ is the $k$-th source of point-to-point
correlated systematic uncertainties in the $i$-th bin,
and $r_k^{(e)}$ the related weight, treated as a free parameter.
In order to fit the $r_k^{(e)}$ values, a penalty must be added
to the definition of the $\chi^2$, as in the second term of
Eq.~(\ref{e.chi2}).

Clearly the evaluation of the multidimensional integrations in
Eqs.~(\ref{e.E}) and (\ref{e.V}) is not practical, especially when
$\mathcal{O}$ is a continuous function such as in the case of FFs.
Instead one can construct an MC representation of
$\mathcal{P}(\bm{a}|{\rm data})$ such that the expectation value
and variance can be evaluated as
\begin{align}
{\rm E}[\mathcal{O}]
&= \frac{1}{n} \sum_{k=1}^{n}\mathcal{O}(\bm{a}_k),
\label{e.Esum}						\\
{\rm V}[\mathcal{O}]
&= \frac{1}{n} \sum_{k=1}^{n}
(\mathcal{O}(\bm{a}_k)-{\rm E}[\mathcal{O}])^2,
\label{e.Vsum}
\end{align}
where the parameters $\{\bm{a}_k\}$ are distributed according to
$\mathcal{P}(\bm{a}|{\rm data})$, and $n$ is the number of points
sampled from the distribution $\mathcal{P}(\bm{a}|{\rm data})$.

Our approach to constructing the Monte Carlo ensemble $\{\bm{a}_k\}$
is schematically illustrated in \fref{workflow}.  The steps in the
IMC procedure can be summarized in the following workflow:
\begin{enumerate}
\item
{\bf Generation of the priors} \\
The \emph{priors} are the initial parameters that are used as
guess parameters for a given least-squares fit.  The resulting
parameters from the fits are called \emph{posteriors}.  During the
initial iteration, a set of priors is generated using a \emph{flat}
sampling in the parameter space.  The sampling region is selected
for the shape parameters $\alpha > -1.9$ and $\beta > 0$, so that
the first moments of all FFs are finite.
The boundary for $\beta$ restricts the distributions to be strictly
zero in the $z \to 1$ limit.  The upper boundaries for $\alpha$ and
$\beta$ are selected to cover typical ranges observed in previous
analysis \cite{HKNS07, DSS07, AKK08}.  Note, however, that the
posteriors can be distributed outside of the initial sampling
region, if this is preferred by the data.

For each subsequent iteration, the priors are generated from a
multivariate Gaussian sampling using the covariance matrix and the
central parameters from the priors of the previous iteration.
The central parameters are chosen to be the median of the priors,
which is found to give better convergence compared with using the
mean.  This sampling procedure further develops the JAM15 methodology
\cite{JAM15}, where the priors were randomly selected from the
previous iteration posteriors.
This allows one to construct priors that are distributed more
uniformly in parameter space as opposed to priors that are
clustered in particular regions of parameter space.  The latter can
potentially bias the results if the number of priors is too small.

\item
{\bf Generation of pseudodata sets} \\
Data resampling is performed by generating pseudodata sets using
Gaussian smearing with the mean and uncertainties of the original
experimental data values.  Each pseudodata point
$\widetilde{{\cal D}}_i$ is computed as
\begin{align}
\widetilde{{\cal D}}_i = {\cal D}_i + R_i\, \alpha_i,
\end{align}
where for each experiment ${\cal D}_i$ and $\alpha_i$ are as in
Eq.~(\ref{e.chi2}), and $R_i$ is a randomly generated number from
a normal distribution of unit width.  A different pseudodata set is
generated for each fit in any given iteration in the IMC procedure.

\item
{\bf Partition of pseudodata sets for cross-validation} \\
To account for possible over-fitting, the cross-validation method
is incorporated.  Each experimental pseudodata set is randomly
divided $50\%/50\%$ into ``training'' and ``validation'' sets.
However, data from any experiment with fewer than 10 points are
not partitioned and are entirely included in the training set.

\item 
{\bf $\chis$ minimization and posterior selection} \\
The $\chi^2$ minimization procedure is performed with the training
pseudodata set using the Levemberg-Marquardt \texttt{lmdiff}
algorithm \cite{More80}.  For every shift in the parameters during
the minimization procedure, the $\chi^2$ values for both training
and validation are computed and stored along with their respective
parameter values, until the best fit for the training set is found.
For each pseudodata set, the parameter vector that minimizes the
$\chi^2$ of the validation is then selected as a posterior.

\item 
{\bf Convergence criterion} \\
The iterative approach of the IMC is similar to the strategy adopted
in the MC VEGAS integration \cite{VEGAS}.  There, one constructs
iteratively a grid over the parameter space such that most of the
sampling is confined to regions where the integrand contributes
the most, a procedure known as \emph{importance sampling}.
Once the grid is prepared, a large amount of samples is generated
until statistical convergence of the integral is achieved.

In Ref.~\cite{JAM15} the convergence of the MC ensemble
$\{ \bm{a}_k \}$ was estimated using the $\chi^2$ distribution.
While such an estimate can give some insight about the convergence
of the posteriors, it is somewhat indirect as it does not involve
the parameters explicitly.  In the present analysis, we instead
estimate the convergence of the eigenvalues of the covariance
matrix computed from the posterior distributions.
To do this we construct a measure given by
\begin{align}
V = \prod_i \sqrt{W_i},
\end{align}
where $W_i$ are the eigenvalues of the covariance matrix.
The quantity $V$ can be interpreted in terms of the hypervolume
in the parameter space that encloses the posteriors, and is
analogous to the ensemble of the most populated grid cells in
a given iteration of the VEGAS algorithm \cite{VEGAS}.
The IMC procedure is then iterated starting from step 1,
until the volume remains unchanged.

\item 
{\bf Generation of the Monte Carlo FF ensemble} \\
When the posteriors volume has reached convergence, a large number
of fits is performed until the mean and expectation
values of the FFs converge.  The goodness-of-fit is then evaluated
by calculating the overall single $\chi^2$ values per experiment
according to
\begin{align}
\chi^2_{(e)}=\sum_i 
\left(
\frac{\mathcal{D}_i^{(e)}-E[T_i^{(e)}]/E[N_i^{(e)}]}{\alpha_i^{(e)}}
\right)^2,
\end{align}
which allows a direct comparison with the original unmodified data.
\end{enumerate}
Finally, note that while the FF parametrization adopted here is
not intrinsically more flexible than in other global analyses,
the MC representation is significantly more versatile and adaptable
in describing the FFs.  Indeed, the resulting averaged central value
of the FFs as a function of $z$ is a linear combination of many
functional shapes, effectively increasing the flexibility of the
parametrization.

\newpage
\section{Data sets}
\label{s.data}

In the current analysis we use all available data sets from
the single-inclusive annihilation process $e^+ e^- \to h X$,
for $h = \pi^\pm$ and $K^\pm$ mesons.
Table~\ref{t:data} summarizes the various SIA experiments,
including the type of observable measured (inclusive or tagged),
center-of-mass energy $Q$, number of data points, and the $\chi^2$
values and fitted normalization factors for each data set.
Specifically, we include data from experiments at
DESY (from the TASSO \cite{TASSO80, TASSO83, TASSO89} and
  ARGUS \cite{ARGUS89} Collaborations),
SLAC (TPC \cite{TPC84, TPC86, TPC88}, HRS \cite{HRS87},
  SLD \cite{SLD04} and BaBar \cite{BaBar13} Collaborations),
CERN (OPAL \cite{OPAL94, OPAL00}, ALEPH \cite{ALEPH95} and
  DELPHI \cite{DELPHI95, DELPHI98} Collaborations) and
KEK (TOPAZ \cite{TOPAZ95} and Belle \cite{Belle13, Leitgab13}
  Collaborations).
Approximately half of the 459\ $\pi^\pm$ data points and 391\
$K^\pm$ data points are near the $Z$-boson pole, $Q \approx M_Z$,
while the most recent, high-precision Belle and BaBar data from
the $B$-factories are at $Q \simeq 10.5$~GeV.
The latter measurements in particular provide a more
comprehensive coverage of the large-$z$ region, and reveal
clearer scaling violation effects compared with the previous
higher-energy measurements.

\begin{table}[t]
\caption{Single-inclusive $e^+ e^-$ annihilation experiments
	used in this analysis, inluding the type of observable
	(inclusive or tagged),
	center-of-mass energy $Q$,
	number of data points $N_{\rm dat}$,
	average fitted correlated normalization
	(when different from ``1''), and
	$\chi^2$ values for pion and kaon production.
	Note that the normalization factors for the
	TASSO data, indicated by $^{(*)}$ in the table, 
        are in the range 0.976 -- 1.184 
	for pions and 0.891 -- 1.033  for kaons.
	For the BaBar pion data \cite{BaBar13} the ``prompt''
	data set is used in the fit discussed in this paper, with 
        normalization and $\chi^2$ values obtained using the 
        ``conventional'' data set in parentheses.}
\scriptsize
\ \\
\input{chi2-tab}

\label{t:data}
\end{table}

In the TPC, OPAL, DELPHI and SLD experiments, light-quark and
heavy-quark events were separated by considering the properties
of final-state hadrons.  In the SLD experiment, for example,
events from the primary $c$ and $b$ quarks were selected by
tracks near the primary interaction point.
For each secondary vertex, the total transverse momentum and
invariant mass were obtained, after which the data were separated
into $c$- and $b$-tagged events depending on the masses and
transverse momenta.  Some events without the secondary vertex
were considered as light-quark ($u,d,s$)-tagged if a track did not
exist with an impact parameter exceeding a certain cutoff value.
Other tagged data sets used different techniques for selecting
the quark-tagged events.
In the OPAL experiment, separated probabilities for $u$, $d$ and $s$
quark fragmentation were also provided, which in practice provide
valuable constraints on the flavor dependence in the light-quark FFs.

For the Belle measurements \cite{Belle13}, the data are provided in
the form $d\sigma^h/dz$, and care must be taken when converting this
to the hadronic FF in Eq.~(\ref{e.empirical}).
The fragmentation energy scale $Q/2$ is reduced by initial-state (ISR)
or final-state (FSR) photon radiation effects, so that the measured
yield involves a variation of this scale.
In practice, the energy scales in the measured events are kept
within 0.5\% of the nominal $Q/2$ value, and an MC simulation
is performed to estimate the fraction of events with ISR or FSR
photon energies $< 0.5\% \times Q/2$.
For each bin the measured yields are reduced by these fractions
to exclude events with large ISR or FSR contributions.
To convert the $d\sigma^h/dz$ data with the ISR/FSR cut to the
total hadronic FF in Eq.~(\ref{e.empirical}) one therefore needs
to correct the theoretical total cross section $\sigma_{\rm tot}$
by multiplying it by the ISR/FSR correction factor, which is
estimated to be 0.64616(3)~\cite{Belle13, Leitgab13}.

For the BaBar experiment \cite{BaBar13}, two data sets were provided,
for ``prompt'' events, which contain primary hadrons or decay products
of lifetimes shorter than $10^{-11}$~s, and ``conventional'' events,
which include decays of lifetimes $(1-3) \times 10^{-11}$~s.
For pions the conventional cross sections are $\sim 5\%-15\%$ larger
than the prompt cross sections, while for kaons these are almost
indistinguishable.  The prompt data are numerically close to the
LEP and SLD measurements after taking into account $Q^2$ evolution,
although the conventional ones are technically closer to most
previous measurements which included all decays.
In our analysis, we consider both data sets, and assess their
impact on the fits phenomenologically.

Finally, our theoretical formalism is based on the fixed-order
perturbation theory, and does not account for resummations of
soft-gluon logarithms or effects beyond the collinear factorization
which may be important at small values of $z$.  To avoid
inconsistencies between the theoretical formalism and the data,
cuts are applied to exclude the small-$z$ region from the analysis.
In practice, we use a cut $z>0.1$ for data at energies below
the $Z$-boson mass and $z>0.05$ for the data at $Q \approx M_Z$.
For kaon data, below $z \approx 0.2$ hadron mass corrections
are believed to play a more prominent role \cite{AAR15}.
Since these are not considered in this analysis, we apply the
cut $z>0.2$ for the low-$Q$ kaon data sets from ARGUS and BaBar.

\section{Analysis results}
\label{s.results}

In this section we present the main results of our IMC analysis.
We first establish the stability of the IMC procedure by examining
specific convergence criteria, and then illustrate the results
for the fragmentation functions through comparisons with data
and previous analyses.
Programs for generating the FFs obtained in this analysis, which
we dub ``JAM16FF'', can be downloaded from Ref.~\cite{JAMgit}.

\subsection{IMC convergence}

We examine two types of convergence tests of the IMC procedure,
namely, the iterative convergence of the priors (the ``grid''),
and the convergence of the final posterior distributions.
As discussed in \sref{imc}, the convergence of the priors can be
tested by observing the variation of the volume $V$ with the number
of iterations, as shown in \fref{vol}.  For each iteration 200 fits
are performed.  During the initial $\sim 10$ iterations, the volume
changes some 9 orders of magnitude, indicating a very rapid variation
of the prior distribution.  After $\sim 30$ iterations, the volume
becomes relatively stable, with statistical fluctuations around 2
orders magnitude due to finite statistics.  The stability of the
prior volume indicates that the region of interest in the parameter
space has been isolated by the IMC procedure.

Having obtained an optimal MC priors sample, a final iteration is
performed with $10^4$ fits.  In \fref{convergence} we illustrate
the statistical properties of the final posterior distribution
by showing averaged ratios of FFs with smaller samples
(100, 200, 500 and 1000) relative to the total $10^4$ samples
(the averaged error bands are displayed only for the 200 and
$10^4$ samples).
Using 200 posterior samples, one obtains uncertainty bands that are
comparable with those with $10^4$ samples.  For the central values
most of the FFs with 200 samples agree well with the $10^4$ samples.
Some exceptions are the $D^\pi_{s^+}$, $D^\pi_g$, $D^K_{d^+}$ and
$D^K_g$ FFs; however, here the differences are in regions where the
FFs are poorly determined and the relative error bands are large.
For practical applications these effects will be irrelevant, and
using a sample of 200 posteriors will be sufficient to give an
accurate representation of FFs.
Unless otherwise stated the results presented in the following
use 200 fits from the final sample.

\subsection{SIA cross sections}

In \fref{chi2} the normalized yields of the final posteriors versus
$\chi^2$ per datum for the training and validation sets are presented
using the full sample of $10^4$ fits.  In the ideal Gaussian limit,
the distributions are expected to peak around 2 \cite{JAM15}.
In practice, inconsistencies between data sets shift the peak
of the distribution to larger $\chi^2/N_{\rm dat}$ values.
This is evident for the pion production case in \fref{chi2},
where the $\chi^2/N_{\rm dat}$ distribution peaks around 2.5.
In contrast, for kaon production the distribution peaks around 2.1.
We stress, however, that even if the peak occurs at 2, it does not
imply consistency among the data sets (or data vs. theory), since
the larger experimental uncertainties in the kaon data sets compared
with the pion can induce such behavior.

\begin{figure}[t]
\includegraphics[width=1\textwidth]{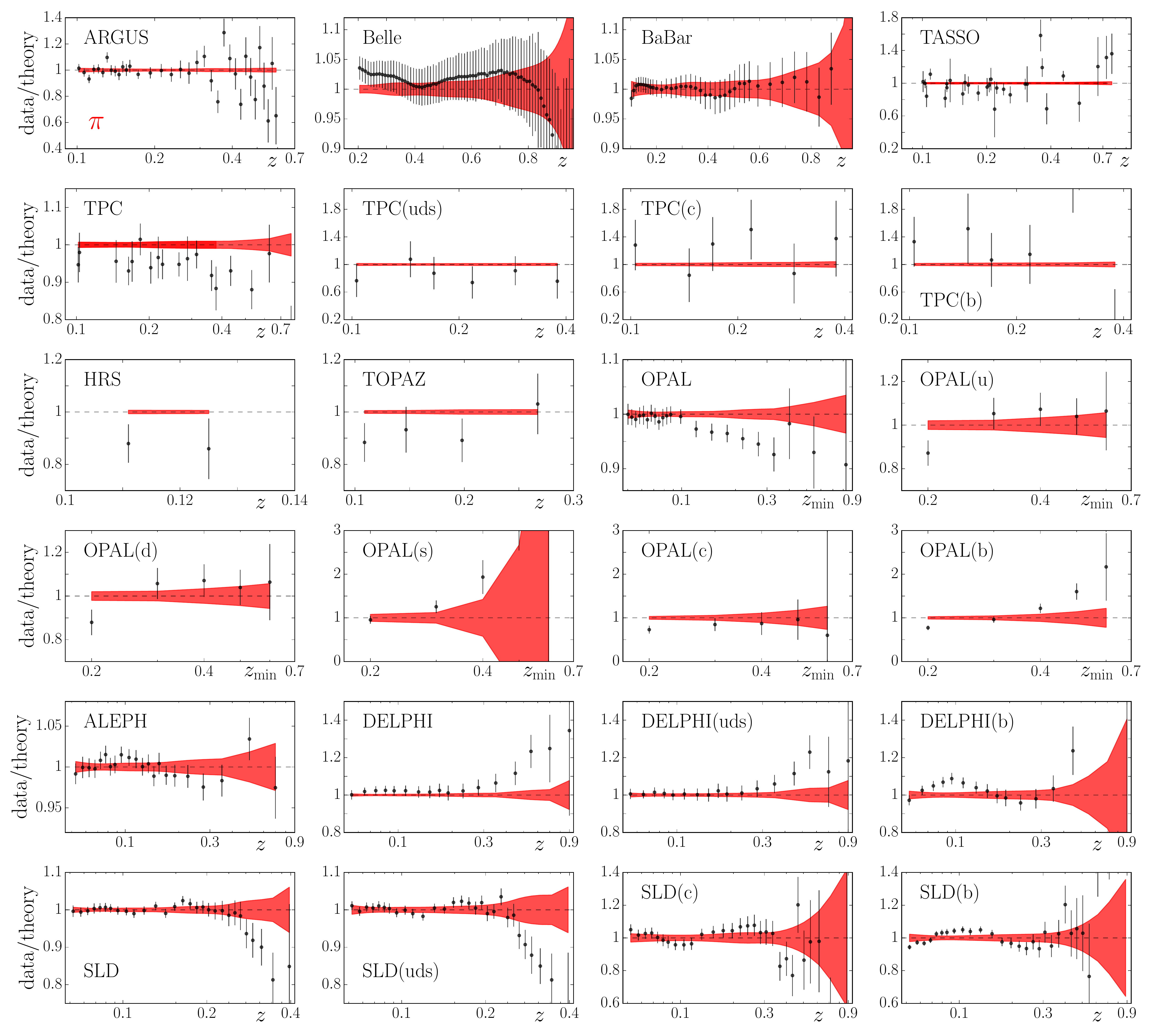}
\caption{Ratio of experimental single-inclusive $e^+ e^-$ cross
	sections to the fitted values versus $z$ (or $z_{\rm min}$
	for OPAL data \cite{OPAL94, OPAL00}) for pion production.
	The experimental uncertainties are indicated by the black
	points, with the fitted uncertainties denoted by the red bands.
	For the BaBar data \cite{BaBar13} the prompt data set is used.}
\label{f.data-thy-pion}
\end{figure}

\begin{figure}[t]
\includegraphics[width=1\textwidth]{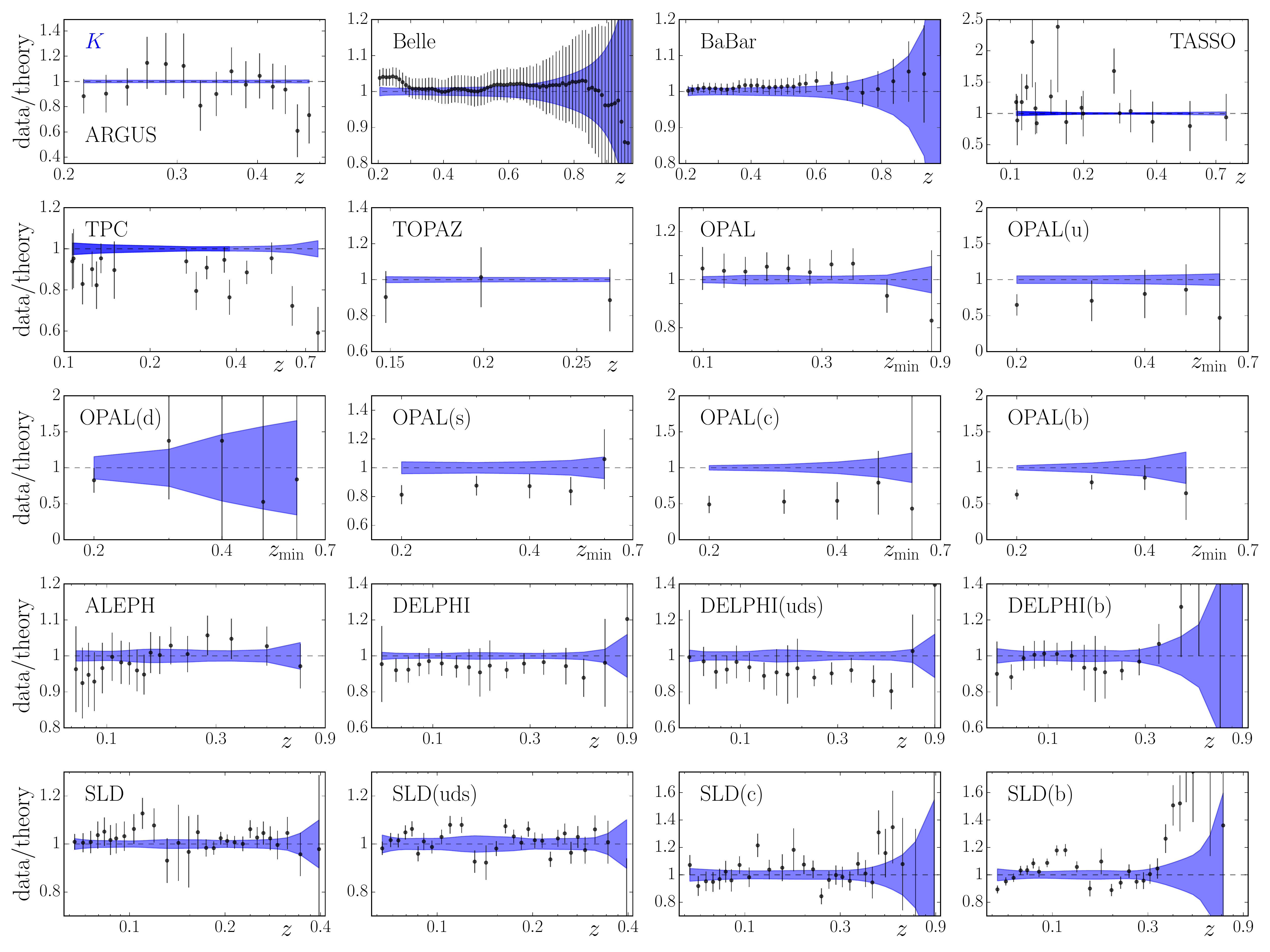}
\caption{As in Fig.~\ref{f.data-thy-pion}, but for kaon production.}
\label{f.data-thy-kaon}
\end{figure}

The ratios of experimental SIA cross sections to the fitted values
are shown in Figs.~\ref{f.data-thy-pion} and \ref{f.data-thy-kaon}
for pions and kaons, respectively.
For the pion production data, at the lower energies $Q \lesssim 30$~GeV
there is good overall agreement between the fitted cross sections
and the data, with the exception of a few sets (TPC, HRS and TOPAZ)
that differ by $\sim 5-10\%$, within relatively large errors.
Interestingly, the older ARGUS data \cite{ARGUS89} are consistent
with the recent high-precision measurements from Belle \cite{Belle13,
Leitgab13} and BaBar \cite{BaBar13}.  We find, however, that the
Belle pion data require an $\approx 10\%$ normalization, which may be
related to the overall normalization correction from initial state
radiation effects \cite{Leitgab13} or other corrections.
This should not, however, affect the $z$ dependence of the
extracted FFs.

A relatively good description is also obtained of the data at higher
energies, $Q = M_Z$, which generally have smaller uncertainties,
although some discrepancies appear at higher $z$ values.
In particular, an inconsistency is apparent between the shapes of
the DELPHI \cite{DELPHI95, DELPHI98} and SLD \cite{SLD04} spectra
at $z \gtrsim 0.4$ for both inclusive and $uds$-tagged data,
with the DELPHI data lying systematically above the fitted results
and SLD data lying below.
For the heavy quark tagged results the agreement with DELPHI and
SLD data is generally better, with only some deviations at the
highest $z$ values where the errors are largest.
The OPAL tagged data \cite{OPAL94, OPAL00} are the only ones that
separate the individual light quark flavors $u$, $d$ and $s$ from
the heavy flavors. 
The latter have rather large $\chi^2$ values for both pion and kaon
data sets, particularly the $b$--tagged sample of the pion case.
While the unfavored $d$--tagged kaon sample is well described in
the fit, the unfavored $s$--tagged pion data appear less consistent
with the theory.
In all cases the OPAL tagged data require a normalization of
$\approx 20\%$.  Note that the observable for the OPAL data is
the $z$-integrated cross section from $z_{\rm min}$ to 1.

The total $\chi^2/N_{\rm dat}$ for the resulting fit to all pion
data sets is $\approx 1.31$.  Using the conventional BaBar pion
data set instead of the prompt gives a slightly worse overall fit,
with $\chi^2/N_{\rm dat} = 1.46$, with the difference coming mostly
from the BaBar and TPC inclusive data sets.
The Belle data, on the other hand, are better fitted when
the conventional BaBar data set is used.
Since the conventional BaBar data lie $\sim 10\%$ higher than
the prompt, which themselves lie slightly below the Belle data,
the Belle cross sections require a normalization shift that is
closer to that needed for the conventional BaBar data.

For the kaon cross sections, the overall agreement between theory and
experiment is slightly better than for pions, mostly because of the
relatively larger uncertainties on the $K$ data.  At low energies,
as was the case for pions, the TPC data \cite{TPC84, TPC86, TPC88}
lie $\approx 10\%$ below the global fit.  Interestingly, though,
the Belle kaon data \cite{Belle13, Leitgab13} do not require as
large a normalization shift as was needed for the Belle pion data
in Fig.~\ref{f.data-thy-pion}.
At energies near the $Z$-boson pole, $Q = M_Z$, the deviations at
large $z$ between the theoretical and experimental cross sections
are not as prominent as for pions, with only the SLD heavy quark
tagged data \cite{SLD04} exhibiting any significant disagreement.
The OPAL flavor-tagged data \cite{OPAL94, OPAL00} generally prefer
an $\approx 10-15\%$ normalization for all quark flavors.
The DELPHI inclusive and light quark tagged data \cite{DELPHI95,
DELPHI98}, which do not include an overall normalization parameter,
appear to systematically lie $\approx 10\%$ below the fitted results
across most of the $z$ range.  Fits to other high energy data sets
generally give good agreement, and the $\chi^2/N_{\rm dat}$ value
for the combined kaon fit is found to be $1.01$.

\subsection{Fragmentation functions}

The fragmentation functions resulting from our IMC analysis are
shown in \fref{FFQ20} at the input scale, which is taken to be
$Q^2=1$~GeV$^2$ for the $u$, $d$, $s$ and $g$ flavors and at the
mass thresholds $Q^2=m_q^2$ for the heavy $c$ and $b$ quarks.
The curve bundles represent random samples of 100 posteriors
from the full set of fitted results, with the central values and
variance bands computed from Eqs.~(\ref{e.E}) and (\ref{e.V})
using the 200 posteriors selected for the final JAM16FF results
\cite{JAMgit}.
Generally the pion FFs have a larger magnitude than the kaon FFs,
with the exception of the strange quark, where the $s^+$ to kaon
distribution $D_{s^+}^{K^+}$ is larger than that for the pion,
$D_{s^+}^{\pi^+}$, over most of the $z$ range.
As expected, the $u^+$ and $d^+$ FFs to $\pi^+$, which correspond
to sums of favored and unfavored distributions and reflect the
valence structure of the pion, are dominant at intermediate and
large values of $z$, $z \gtrsim 0.2$.
In the limit of exact isospin symmetry (neglecting mass differences
between $u$ and $d$ quarks), these are in fact identical,
$D_{u^+}^{\pi^+} = D_{d^+}^{\pi^+}$.
The $s^+$ to pion distribution, in contrast, is smaller in magnitude,
with a peak value at $x \sim 0.3-0.4$ that is less than $\approx 1/2$
that for the nonstrange quarks.  Since this distribution is entirely
unfavored, and constrained mostly by the scant flavor-tagged data,
it has somewhat larger uncertainties compared with the nonstrange FFs.
Note that our analysis uses two shapes for the favored
$D_{u^+}^{\pi^+} = D_{d^+}^{\pi^+}$ FFs, but one shape for all other
pion distributions, Eqs.~(\ref{e.piparam}).

For the heavy quark FFs to pions, the characteristic differences
between the $s^+$, $c^+$ and $b^+$ distributions generally reflect
the different masses of the quarks, with larger mass corresponding
to softer distributions.  The $c^+$ and $b^+$ FFs, in particular, are
large at low $z$ values, $z \lesssim 0.1$, and comparable to the
light-quark FFs evolved to the same scale.  (Note that the heavy
quark distributions exist only above the mass threshold, $Q > m_q$.)
The gluon FF $D_g^{\pi^+}$ is less singular, but is strongly peaked
at $z \approx 0.25$ at the input scale.  Its uncertainties are also
larger than those for the favored distributions, as their effects
on the SIA cross sections are of higher order in $\alpha_s$.

For the fragmentation to kaons, one of the most conspicuous
differences with pions is the large magnitude of the strange FF
$D_{s^+}^{K^+}$ at intermediate and high values of $z$, where
it is comparable to the $u^+$ and $d^+$ FFs to pions.
Reflecting the valence quark structure of $K^\pm$, the $D_{u^+}^{K^+}$
FF is also similar in size, but because of the mass difference between
the strange and nonstrange light quarks there is no reason for the
favored $u^+$ and $s^+$ fragmentation to kaons to be equal.
In fact, we find $D_{s^+}^{K^+} \gtrsim D_{u^+}^{K^+}$ at high
values of $z$.  Unlike for pions, the $d^+$ fragmentation to kaons
is unfavored, $D_{d^+}^{K^+} \ll D_{u^+}^{K^+}$, with relatively
large uncertainties, peaking at $z \sim 0.1$ and playing a role
similar to that of $s^+$ fragmentation to $\pi^+$.
The heavy quark FFs to kaons are also sizeable compared with the
light quark functions, but peak at slightly larger $z$ values than
the corresponding pion FFs.  The gluon FF to kaons, $D_g^{K^+}$,
peaks at rather high $z$ values, $z \approx 0.85$, at the input
scale, consistent with the findings of some earlier analyses
\cite{HKNS07}, and is very small in magnitude.

\begin{figure}[t]
\includegraphics[width=1\textwidth]{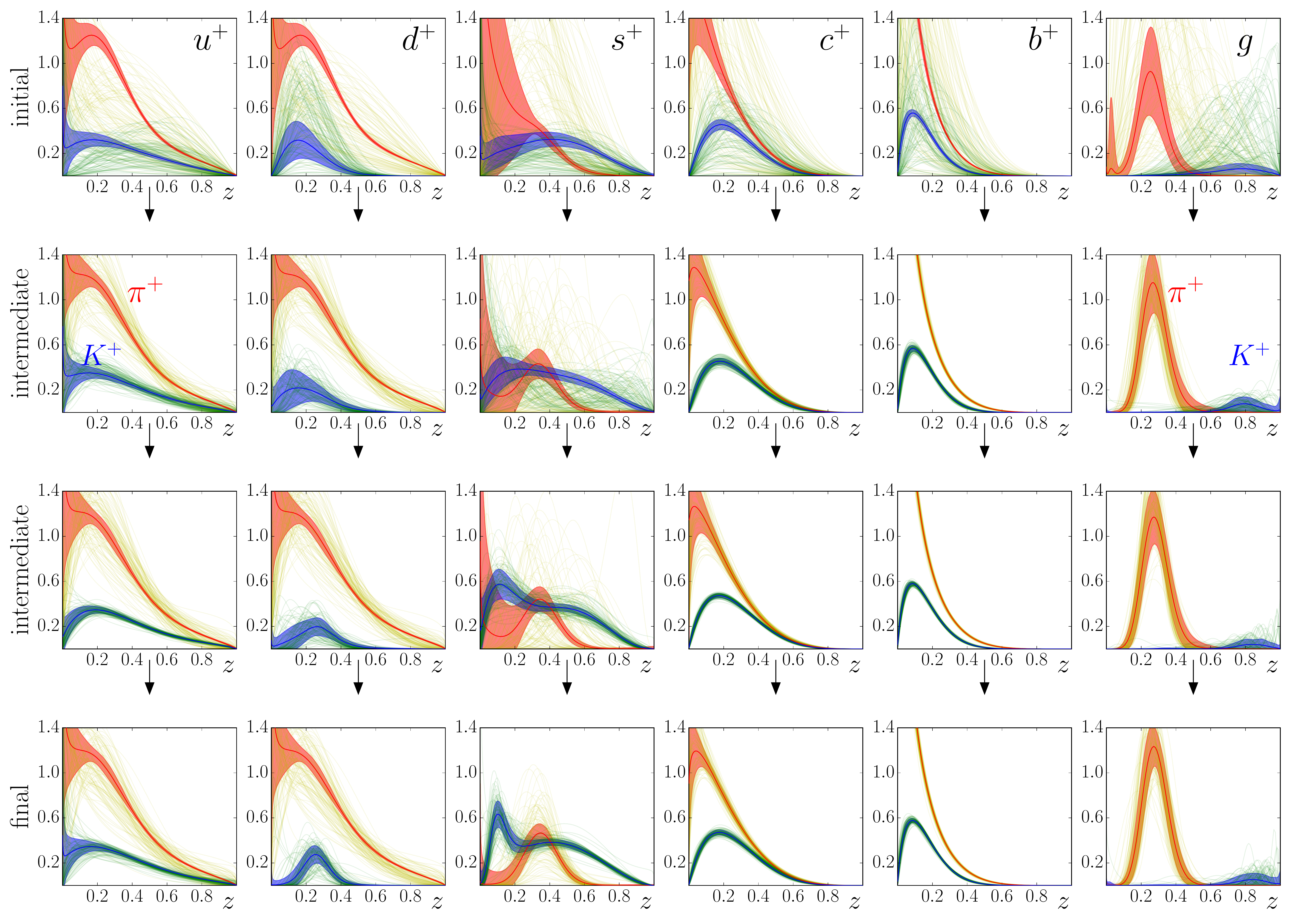}
\caption{Iterative convergence of the $\pi^+$ (red bands)
	and $K^+$ (blue bands) fragmentation functions for the
	$u^+$, $d^+$, $s^+$, $c^+$, $b^+$ and $g$ flavors
	(in individual columns) at the input scale.
 	The first row shows the initial flat priors (single
	yellow curves for $\pi^+$ and green curves for $K^+$)
	and their corresponding posteriors (error bands).
	The second and third row are selected intermediate
	snapshots of the IMC chain, and the last row shows
	the priors and posteriors of the final IMC iteration.}
\label{f.IMC}
\end{figure}

The unusual shapes of some of the FFs, such as the gluon to $\pi^+$
and $K^+$ or the unfavored light quark FFs, lead to the natural
question of whether these are robust distributions or possibly
artifacts of the fitting procedure.
We can address this by observing snapshots in the IMC chain,
as illustrated in Fig.~\ref{f.IMC}, where the FFs from selected
iteration steps are plotted at the input scale as a function of $z$.
The first and last rows in Fig.~\ref{f.IMC} show the initial
and final steps in the IMC procedure, respectively.
In addition to the posterior shapes and uncertainties, we display
in each row the prior distributions as individual curves.
After performing the initial iteration, the large spread in the
prior FFs due to the flat sampling of the parameter space is
reduced significantly, especially for distributions that are
more strongly constrained by the SIA data.
For the FFs that are less directly constrained by the data,
more iterations are needed before convergence is reached, as
illustrated by the $s^+$ to $\pi^+$ distribution, for example.
We find that after $\approx 30$ iterations all of the distributions
become stable, which is consistent with the convergence of the
volumes observed in \fref{vol}.

Although the peaks in some of the FFs, such as $D_g^{\pi^+, K^+}$
and $D_{s^+}^{\pi^+}$, are prominent at the input scale, after $Q^2$
evolution these become largely washed out.  This is illustrated in
\fref{FFevo}, where the FFs are evolved to a common scale for all
FFs that are above the quark threshold, namely at $Q^2=1, 10$ and
100~GeV$^2$ and at the $Z$-boson pole, $Q^2 = M_Z^2$.  Recall that
the lowest $Q^2$ in any of the data sets is $\approx 100$~GeV$^2$,
so the shapes at $Q^2=1$ and 10~GeV$^2$ are not directly compared
with experimental data and are shown for illustration only.

\begin{figure}[t]
\includegraphics[width=1\textwidth]{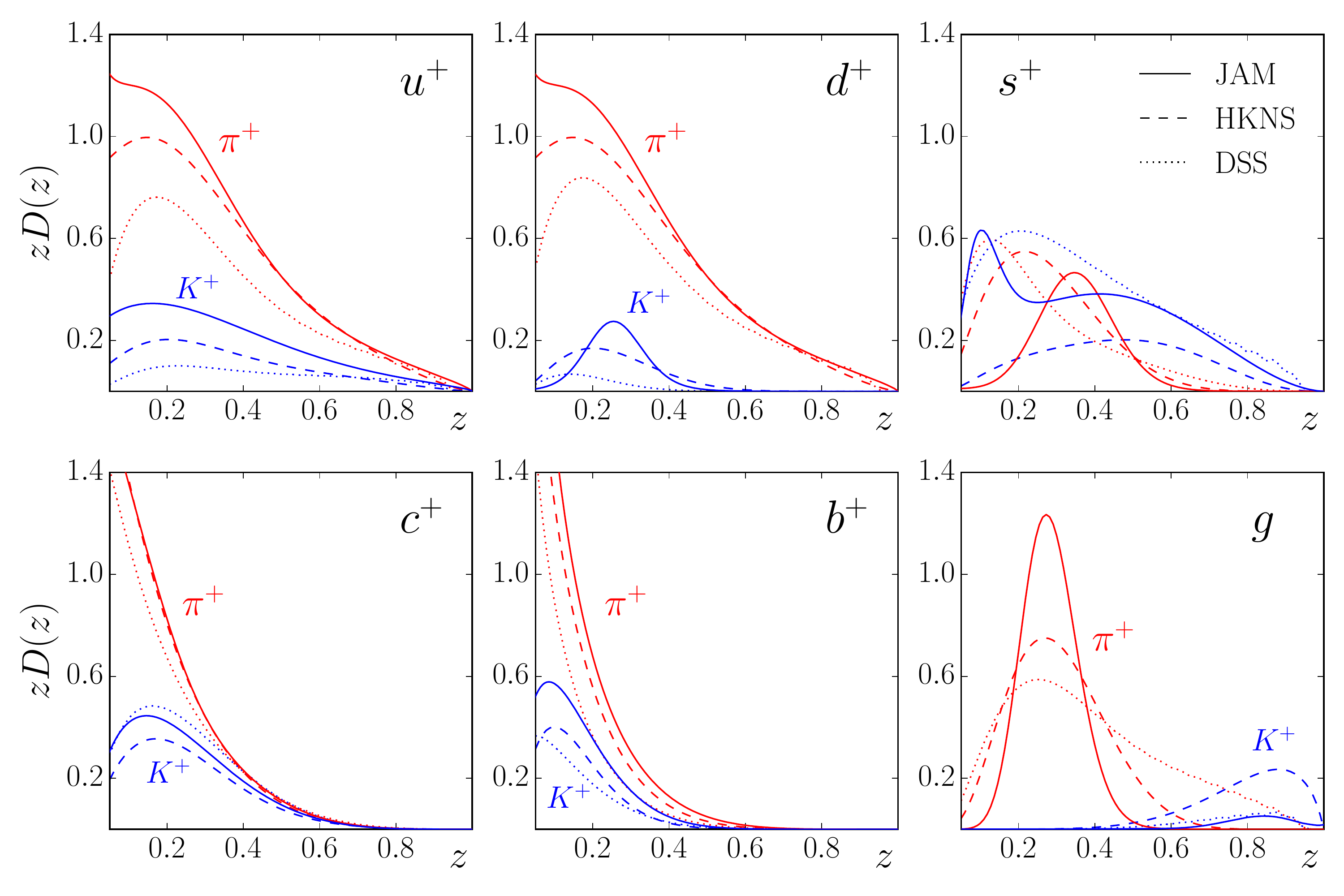}
\caption{Comparison of the JAM fragmentation functions (solid curves)
	for $\pi^+$ (red curves) and $K^+$ (blue curves) with the
	HKNS \cite{HKNS07} (dashed curves) and DSS \cite{DSS07}
	(dotted curves) parametrizations at the input scale
	$Q^2=1$~GeV$^2$ for the light quark and gluon distributions,
	and $Q^2=10$ and 20~GeV$^2$ for the $c^+$ and $b^+$ flavors,
	respectively.}
\label{f.comparisonI}
\end{figure}

Compared with parametrizations from other global FF analyses,
our fitted FFs are qualitatively similar for the most part,
but reveal important differences for specific distributions,
as Fig.~\ref{f.comparisonI} illustrates.  For pions, our $u^+$
and $d^+$ distributions are close to the HKNS \cite{HKNS07} and
DSS \cite{DSS07} results at large $z$, but are $\sim 20\%-30\%$
larger in magnitude at low $z$ values, $z \lesssim 0.3$.
The strange quark to pion FF peaks at somewhat larger $z$ than
the nonstrange, with a magnitude similar to that in previous fits.
The peak in the gluon FF at $z \approx 0.25$ coincides with that
in the HKNS and DSS gluon FFs, but our distribution is rather
more narrow with a smaller large-$z$ tail.

The comparison between the various parametrizations for the kaon FFs
is quite instructive, especially for the light quark flavors and the
gluon.  The favored $D_{u^+}^{K^+}$ and $D_{s^+}^{K^+}$ FFs in our
fit turn out to be of comparable magnitude, with the $u^+$ closer
to the HKNS results and $s^+$ closer to DSS.  In particular, for
the $u$ quark to kaon FF our result is $\approx 30\%-50\%$ larger
than HKNS, but some 2--3 times greater than DSS over the range
$0.2 \lesssim z \lesssim 0.9$.
On the other hand, the strange to kaon FF lies between the HKNS
and DSS results at intermediate $z$ values, but coincides with
the DSS at $z \gtrsim 0.5$.
Interestingly, we do not observe the large excess of $s$ to $K$
fragmentation over $u$ to $K$ found in the DSS analysis, which
has important phenomenological consequences for the extraction
of the polarized strange quark PDF from semi-inclusive DIS data
\cite{LSS11, LSS15}.

Recall that in our analysis we use two shapes for the favored
$D_{u^+}^{K^+}$ and $D_{s^+}^{K^+}$ FFs, Eqs.~(\ref{e.Kparam}),
and one shape for all other kaon distributions.
In contrast, previous analyses \cite{HKNS07, DSS07} parametrized
the $u^+$ and (the unfavored) $\bar u$ functions separately,
assuming that at the input scale
	$D_{\bar u}^{K^+} = D_d^{K^+} = D_{\bar d}^{K^+}$.
In contrast, with the IMC procedure in the present analysis we
do not impose any relation between the $\bar u$ and $\bar d$ FFs,
parametrizing only the $q^+$ distributions as constrained by data.

For the gluon to kaon FF we find a similarly hard distribution as in
earlier analyses, peaking at rather large $z$ values, $z \sim 0.8$
at the input scale.  The harder shape of $D_g^{K^+}$ compared with
$D_g^{\pi^+}$ can be understood in terms of the higher energy needed
for a gluon to split to an $s \bar s$ pair than to a $u \bar u$ or
$d \bar d$ pair in the pion case \cite{HKNS07}.

\begin{figure}[t]
\includegraphics[width=1\textwidth]{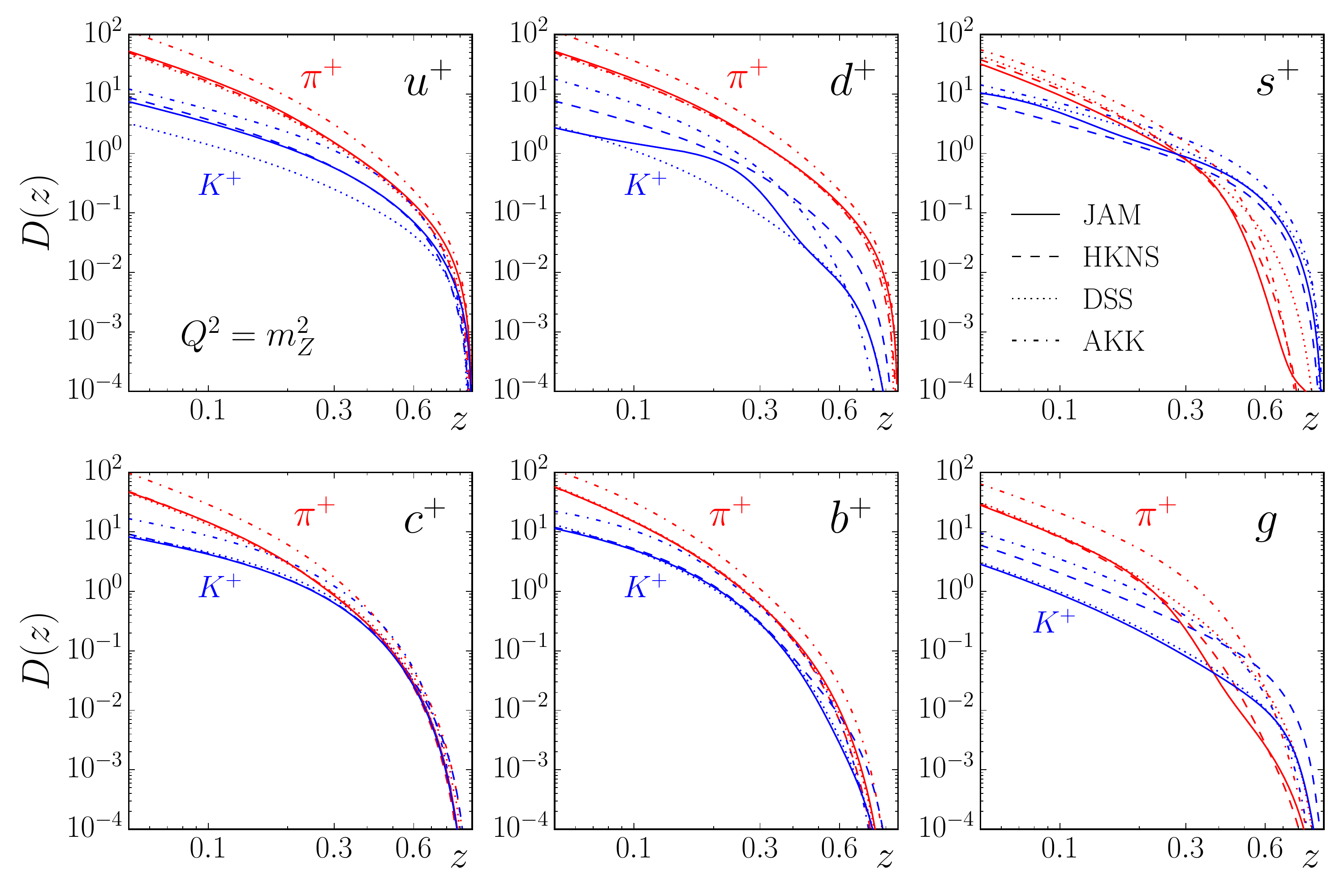}
\caption{Comparison of the JAM fragmentation functions (solid curves)
	for $\pi^+$ (red curves) and $K^+$ (blue curves) with the
	HKNS \cite{HKNS07} (dashed curves), DSS \cite{DSS07}
	(dotted curves) and AKK \cite{AKK08} (dot-dashed curves)
	evolved to a common scale $Q^2=M_Z^2$.  Note that the
	fragmentation functions $D(z)$ are shown rather than
	$z D(z)$.}
\label{f.comparisonII}
\end{figure}

Despite the striking shape of the gluon FF at the input scale,
it is almost entirely washed out after $Q^2$ evolution to the
$Z$-boson scale, as Fig.~\ref{f.comparisonII} illustrates.
Here the FFs $D(z)$ (rather than $z D(z)$) are compared for the
HKNS \cite{HKNS07}, DSS \cite{DSS07} and the AKK \cite{AKK08}
parametrizations.
Viewed on a logarithmic scale, the qualitative features of the
shapes of FFs are similar across all the parametrizations,
especially the HKNS, DSS and the present fit.  The AKK results
generally lie above the other parametrizations in the low-$z$
region, while more variation is observed at higher $z$ values.

\section{Conclusion}
\label{s.conclusion}

We have performed the first Monte Carlo based QCD analysis for parton
to hadron fragmentation functions within collinear factorization,
using all existing single-inlusive $e^+ e^-$ annihilation data into
pions and kaons.  In particular, we include the recent high-precision
SIA data from the Belle \cite{Belle13, Leitgab13} and BaBar
\cite{BaBar13} Collaborations, which significantly extend the
kinematical coverage to large values of $z$.

Our analysis is based on the iterative Monte Carlo approach, first
adopted in the recent QCD analysis of polarized PDFs \cite{JAM15},
which provides a robust determination of expectation values and
uncertainties for the FFs.  We further extended this methodology by
sampling new priors from previous iterations using a multivariate
Gaussian distribution, implementing a new strategy for assessing
the convergence of the IMC chain by considering the covariance
matrix of the posterior distributions.  This allowed us to sample
fairly the parameter eigenspace after each iteration instead of
the posteriors, which can exhibit several distinct solutions.
We find that an accurate representation of the FFs can be
attained with a sample of 200 fits.

We obtained a relatively good overall description of the pion
and kaon SIA data at both low and high center-of-mass energies,
despite some tensions between the high-energy DELPHI and SLD pion
data sets in the large-$z$ region.  For the kaon data a very good
$\chi^2/N_{\rm dat} \sim 1$ was achieved, partly because of larger
experimental uncertainties and also less evident tensions between
data sets.

The shapes of the pion FFs are qualitatively similar to those
found in previous analyses, with the exception of the unfavored
$D_{s^+}^{\pi^+}$ and the $D_g^{\pi^+}$ distributions.
The latter is more strongly peaked around the maximum at
small $z$ values than either the HKNS or DSS results,
while the former has a somewhat harder $z$ distribution.
The kaon FFs, on the other hand, show greater deviation from
the earlier results.  Here, the favored $D_{s^+}^{K^+}$ function
is similar in magnitude to that from the DSS parametrization
\cite{DSS07} for $0.5 \lesssim z \lesssim 1$, but displays
important differences at $z \lesssim 0.5$ that stem from the
greater flexibility of the parametrization used in our analysis.
We also find a larger magnitude of the $D_{u^+}^{K^+}$ FF at
moderate to low $z$ values compared with the DSS fit in particular.
In contrast, the gluon to kaon distribution, which peaks at very
large $z$ values, $z \sim 0.85$, but with a very small magnitude,
is consistent with the DSS result.  The disparity between the
fitted $D_g^{\pi^+}$ and $D_g^{K^+}$ functions is particularly
striking.
At energies on the order of the $Z$-boson mass, the evolved
distributions are much more similar to those of the previous
analyses, with the exception of the $D_g^{\pi^+}$ and
$D_{s^+}^{\pi^+}$ FFs.

The partial separation of the FFs for the various quark flavors
has been possible because of the existence of the tagged flavor
data and the $Q^2$ dependence of SIA cross sections, from low
$Q \sim 10$~GeV up to the $Z$-boson mass, selecting differently
weighted combinations of FFs in the $\gamma$ and $Z$-exchange
cross sections.
To further decompose the quark and antiquark FFs, and better
constrain the gluon fragmentation, additional information will
be needed from SIDIS and meson production in $pp$ collisions.
More immediately, it will be particularly interesting to examine
the effect of the strange to kaon fragmentation on the extraction
of the polarized strange quark PDF $\Delta s$ from SIDIS data.
A combined analysis of polarized DIS and SIDIS data and SIA
cross sections is currently in progress \cite{SIDIS16}.

\section*{Acknowledgment}

We are grateful to Hrayr Matevosyan for helpful discussions.
This work was supported by the US Department of Energy (DOE)
contract No.~DE-AC05-06OR23177, under which Jefferson Science
Associates, LLC operates Jefferson Lab, and by the DOE contract
DE-SC008791.  N.S. thanks KEK and J-PARC for their hospitality
during a visit where some of this work was performed.  The work of
S.K. and M.H was supported by JSPS KAKENHI Grant Number JP25105010.

\appendix
\section{Hard scattering coefficients}

For completeness, in this appendix we give the hard coefficient
functions in Mellin moment space at NLO.  For the quark case,
the NLO coefficient is \cite{Kretzer00, GRV90}
\begin{eqnarray}
\widetilde{\mell{H}}^{(1)}_q(N,Q^2,\mu_R^2=Q^2,\muFF^2)
&=& 2\, C_F
\biggr[
  5 S_2(N)
+ S_1^2(N) 
+ S_1(N) \left( \frac{3}{2} - \frac{1}{N(N+1)} \right)
- \frac{2}{N^2}				\nonumber\\
& & \hspace*{1cm}
+ \frac{3}{(N+1)^2}
- \frac{3}{2} \frac{1}{(N+1)}
- \frac{9}{2} + \frac{1}{N}            \nonumber\\
& & \hspace*{1cm}
+ \left( \frac{1}{N(N+1)} - 2 S_1(N) + \frac{3}{2} \right)
  \ln\frac{Q^2}{\muFF^2}
\biggr],
\end{eqnarray}
while for the gluon one has
\begin{eqnarray}
\widetilde{\mell{H}}^{(1)}_g(N,Q^2,\mu_R^2=Q^2,\muFF^2)
&=& 4\, C_F
\biggr[
- S_1(N) \frac{N^2+N+2}{(N-1) N (N+1)}
- \frac{4}{(N-1)^2}
+ \frac{4}{N^2}					\nonumber\\
& & \hspace*{0cm}
- \frac{3}{(N+1)^2}
+ \frac{4}{(N-1)N}
+ \frac{N^2 + N + 2}{N(N^2-1)}
  \ln\frac{Q^2}{\muFF^2}
\biggr],
\end{eqnarray}
where $C_F = 4/3$.
%
%
Here the harmonic sums $S_1(N)$ and $S_2(N)$ can be written in terms
of the Euler-Mascheroni constant $\gamma_E$, the polygamma function
$\psi_N$, and the Riemann zeta function $\zeta$, analytically
continued to complex values of $N$ \cite{GRV90},
\begin{eqnarray}
S_1(N) &=& \sum_{j=1}^N \frac{1}{j}\
\longrightarrow\ \gamma_E + \psi^{(0)}_{N+1},	\\
S_2(N) &=& \sum_{j=1}^N \frac{1}{j^2}\
\longrightarrow\ \zeta(2) - \psi^{(1)}_{N+1},
\end{eqnarray}
where the $m$-th derivative of the polygamma function $\psi^{(m)}_N$
is given by
\begin{eqnarray}
\psi^{(m)}_N
&=& \frac{d^m \psi_N}{dN^m}\
 =\ \frac{d^{m+1} \ln\Gamma(N)}{dN^{m+1}}.
\end{eqnarray}

\newpage
\section{Timelike splitting functions}

The $N$-th moments of the splitting functions in the timelike region,
up to ${\cal O}(a_s^3)$ corrections, can be written for the general
case when $\muR \neq \muFF$ as \cite{PEGASUS}
\begin{eqnarray}
\mell{P}_{ij}(N,\muR^2,\muFF^2)
&=& a_s(\muR^2)\,
    \mell{P}_{ij}^{(0)}(N)\,
+\, a_s^2(\muR^2)\,
    \Big( \mell{P}_{ij}^{(1)}(N)
	- \beta_0 \mell{P}_{\rm NS}^{(0)}(N) \ln\frac{\muFF^2}{\muR^2}
    \Big),
\end{eqnarray}
where $\beta_0 = 11-2 n_f/3$.
At leading order the timelike splitting function moments are given
by the well-known expressions \cite{Floratos81, Weigl96}
\begin{subequations}
\begin{eqnarray}
\mell{P}^{(0)}_{\rm NS^\pm}
&=& \mell{P}^{(0)}_{qq}\
 =\ -C_F \left[ 4 S_1(N) - 3 - \frac{2}{N(N+1)} \right],
\label{e.P0NS}							\\
\mell{P}^{(0)}_{qg}
&=& 4 n_f C_F \frac{N^2 + N + 2}{N(N-1)(N+1)},
\label{e.P0qg}							\\
\mell{P}^{(0)}_{gq}
&=& \frac{N^2 + N + 2}{N(N+1)(N+2)},
\label{e.P0gq}							\\
\mell{P}^{(0)}_{gg}
&=& -C_A \left[ 4 S_1(N) - \frac{11}{3} - \frac{4}{N(N-1)}
		- \frac{4}{(N+1)(N+2)}
	 \right] - \frac{2n_f}{3},
\label{e.P0gg}
\end{eqnarray}
\end{subequations}
where $C_A = 3$.
Note that our notation for the off-diagonal timelike splitting
functions $\mell{P}^{(0)}_{qg}$ and $\mell{P}^{(0)}_{gq}$ is
opposite to that in Ref.~\cite{GRV93}.

At NLO accuracy, the timelike splitting function moments are
given by \cite{Floratos81, GRV93, Curci80}
\begin{subequations}
\label{e.P1}
\begin{eqnarray}
\mell{P}^{(1)}_{\rm NS^\pm}
&=& -C^2_F
    \left[
    8 S_1(N) \frac{(2N+1)}{N^2 (N+1)^2}
    + 8 \left( 2 S_1(N) - \frac{1}{N(N+1)} \right)
	\big( S_2(N) - S_{2\pm}'(\tfrac{N}{2}) \big)
    \right.						\nonumber\\
& & \left.
    + 12 S_2(N) + 32 \widetilde{S}_\pm(N)
    - 4 S_{3\pm}'(\tfrac{N}{2})
    - \frac{3}{2}
    - 4 \frac{(3N^3+N^2-1)}{N^3(N+1)^3}
    \mp 8 \frac{(2N^2+2N+1)}{N^3(N+1)^3}
    \right]						\nonumber\\
&-& C_A C_F
    \left[
    \frac{268}{9} S_1(N)
    - 4 \left( 2 S_1(N)- \frac{1}{N(N+1)} \right)
	\big( 2S_2(N) - S_{2\pm}'(\tfrac{N}{2}) \big)
    - \frac{44}{3} S_2(N)
    - \frac{17}{6}
    \right.						\nonumber\\
& & \left.
    - 16 \widetilde{S}_\pm(N)
    + 2 S_{3\pm}'(\tfrac{N}{2}) 
    - \frac29 \frac{(151N^4+236N^3+88N^2+3N+18)}{N^3(N+1)^3}
    \pm 4 \frac{(2N^2+2N+1)}{N^3(N+1)^3}
    \right]						\nonumber\\
&-& \frac{1}{2} n_f C_F
    \left[
    - \frac{80}{9} S_1(N)
    + \frac{16}{3} S_2(N)
    + \frac23
    + \frac{8}{9} \frac{(11N^2+5N-3)}{(N^2(N+1)^2)}
    \right]\
 +\ \Delta_{\rm NS}^{(1)},
\label{e.P1NS}						\\
& &							\nonumber\\
& &							\nonumber\\
\mell{P}^{(1)}_{qq}
&=& \mell{P}^{(1)}_{\rm NS^+}\
 +\ n_f C_F
    \left[
      \frac{(5N^5+32N^4+49N^3+38N^2+28N+8)}{(N-1)N^3(N+1)^3(N+2)^2}
    \right]
 +\ \Delta_{qq}^{(1)},
\label{e.P1qq}						\\
& &							\nonumber\\
& &							\nonumber\\
\mell{P}^{(1)}_{gg}
&=& -\frac12 n_f C_A
    \left[
    - \frac{80}{9} S_1(N)
    + \frac{16}{3}
    + \frac{8}{9}
      \frac{(38N^4+76N^3+94N^2+56N+12)}{(N-1)N^2(N+1)^2(N+2)}
    \right]						\nonumber\\
&-& \frac12 n_f C_F
    \left[
    4 + 8 \frac{(2N^6+4N^5+N^4-10N^3-5N^2-4N-4)}{(N-1)N^3(N+1)^3(N+2)}
    \right]						\nonumber\\
&-& C^2_A
    \left[
    \frac{268}{9} S_1(N)
    + 32 S_1(N)
      \frac{(2N^5+5N^4+8N^3+7N^2-2N-2)}{(N-1)^2 N^2(N+1)^2(N+2)^2}
    - \frac{32}{3}
    \right.						\nonumber\\
& & \hspace*{0.7cm}
    + 16 S_{2+}'(\tfrac{N}{2})
      \frac{(N^2+N+1)}{(N-1)N(N+1)(N+2)}
    - 8 S_1(N) S_{2+}'(\tfrac{N}{2})
    + 16 \widetilde{S}_+(N)
    - 2 S_{3+}'(\tfrac{N}{2})				\nonumber\\
& & \hspace*{0.7cm}
    - \frac29
      \frac{(457N^9+2742N^8+6040N^7+6098N^6+1567N^5-2344N^4-1632N^3)}
           {(N-1)^2 N^3(N+1)^3(N+2)^3}			\nonumber\\
& & \left. \hspace*{0.7cm}
    - \frac29
      \frac{(560N^2+1488N+576)}
	   {(N-1)^2 N^3(N+1)^3(N+2)^3}
    \right]
 +\ \Delta_{gg}^{(1)},
\label{e.P1gg}
\end{eqnarray}
\begin{eqnarray}
\mell{P}^{(1)}_{qg}
&=& 2 n_f C_F^2
    \left[
      \left( S_1^2(N) - 3 S_2(N) - \frac{2\pi^2}{3} \right)
      \frac{(N^2+N+2)}{(N-1)N(N+1)}
    \right.						\nonumber\\
& & \hspace*{0.7cm}
    + 2 S_1(N)
      \left(
        \frac{4}{(N-1)^2}
      - \frac{2}{(N-1)N}
      - \frac{4}{N^2}
      + \frac{3}{(N+1)^2}
      - \frac{1}{(N+1)}
      \right)
    - \frac{8}{(N-1)^2 N}				\nonumber\\
& & \left. \hspace*{0.7cm}
    + \frac{8}{(N-1) N^2}
    + \frac{2}{N^3}
    + \frac{8}{N^2}
    - \frac{1}{2N}
    + \frac{1}{(N+1)^3}
    - \frac{5}{2(N+1)^2}
    + \frac{9}{2(N+1)}
    \right]						\nonumber\\
&+& 2 n_f C_F C_A
    \left[
      \left(
      - S_1^2(N) + 5 S_2(N) - G^{(1)}(N) + \frac{\pi^2}{6}
      \right)
      \frac{(N^2+N+2)}{(N-1)N(N+1)}
    \right.						\nonumber\\
& & \hspace*{0.7cm}
    + 2 S_1(N)
      \left(
      - \frac{2}{(N-1)^2} + \frac{2}{(N-1)N} + \frac{2}{N^2}
      - \frac{2}{(N+1)^2} + \frac{1}{N+1}	
      \right)						\nonumber\\
& & \hspace*{0.7cm}
    - \frac{8}{(N-1)^3}
    + \frac{6}{(N-1)^2}
    + \frac{17}{9(N-1)}
    + \frac{4}{(N-1)^2 N}
    - \frac{12}{(N-1)N^2}
    - \frac{8}{N^2}
    + \frac{5}{N}					\nonumber\\
& & \left. \hspace*{0.7cm}
    - \frac{2}{N^2(N+1)}
    - \frac{2}{(N+1)^3}
    - \frac{7}{(N+1)^2}
    - \frac{1}{N+1}
    - \frac{8}{3(N+2)^2}
    + \frac{44}{9(N+2)}
    \right],
\label{e.P1qg}                                          \nonumber\\
& &							\\
& &                                                     \nonumber\\
\mell{P}^{(1)}_{gq}
&=& \frac13 n_f
    \left[
      S_1(N+1) \frac{(N^2+N+2)}{N(N+1)(N+2)}
    + \frac{1}{N^2}
    - \frac{5}{3N}
    - \frac{1}{N(N+1)}
    - \frac{2}{(N+1)^2}
    \right.						\nonumber\\
& & \left. \hspace*{0.7cm}
    + \frac{4}{3(N+1)}
    + \frac{4}{(N+2)^2}
    - \frac{4}{3(N+2)}
    \right]						\nonumber\\
&+& \frac14 C_F
    \left[
      \Big( - 2 S_1^2(N+1) + 2 S_1(N+1) + 10 S_2(N+1) \Big)
      \frac{(N^2+N+2)}{N(N+1)(N+2)}
    \right.						\nonumber\\
& & \hspace*{0.7cm}
    + 4 S_1(N+1)
      \left(
      - \frac{1}{N^2}
      + \frac{1}{N}
      + \frac{1}{N(N+1)}
      + \frac{2}{(N+1)^2}
      - \frac{4}{(N+2)^2}
      \right)						\nonumber\\
& & \hspace*{0.7cm}
    - \frac{2}{N^3}
    + \frac{5}{N^2}
    - \frac{12}{N}
    + \frac{4}{N^2(N+1)}
    - \frac{12}{N(N+1)^2}
    - \frac{6}{N(N+1)}					\nonumber\\
& & \left. \hspace*{0.7cm}
    + \frac{4}{(N+1)^3}
    - \frac{4}{(N+1)^2}
    + \frac{23}{N+1}
    - \frac{20}{N+2}
    \right]						\nonumber\\
&+& \frac14 C_A
    \left[
      \left(
        2 S_1^2(N+1)
      - \frac{10}{3} S_1(N+1)
      - 6 S_2(N+1)
      + 2 G^{(1)}(N+1)
      - \pi^2
      \right)
    \right.						\nonumber\\
& & \hspace*{1.5cm}
    \times \frac{(N^2+N+2)}{N(N+1)(N+2)}		\nonumber\\
& & \hspace*{0.7cm}
    - 4 S_1(N+1)
      \left(
      - \frac{2}{N^2}
      + \frac{1}{N}
      + \frac{1}{N(N+1)}
      + \frac{4}{(N+1)^2}
      - \frac{6}{(N+2)^2}
      \right)						\nonumber\\
& & \hspace*{0.7cm}
    - \frac{40}{9(N-1)}
    + \frac{4}{N^3}
    + \frac{8}{3N^2}					
    + \frac{26}{9N}
    - \frac{8}{N^2 (N+1)^2}
    + \frac{22}{3N(N+1)}
    + \frac{16}{(N+1)^3}				\nonumber\\
& & \left. \hspace*{0.7cm}
    + \frac{68}{3(N+1)^2}
    - \frac{190}{9(N+1)}
    + \frac{8}{(N+1)^2(N+2)}
    - \frac{4}{(N+2)^2}
    + \frac{356}{9(N+2)}
   \right],
\label{e.P1gq}
\end{eqnarray}
\end{subequations}
where the terms
\begin{subequations}
\begin{eqnarray}
\Delta_{\rm NS}^{(1)}
&=& C_F^2
    \left[ -4 S_1(N) + 3 + \frac{2}{N(N+1)}
    \right]
    \left[ 2 S_2(N) - \frac{\pi^2}{3} - \frac{2N+1}{N^2(N+1)^2}
    \right],						\\
& &                                                     \nonumber\\
\Delta_{qq}^{(1)}
&=& \frac12 n_f C_F
    \left[
    - \frac{80}{9}\frac{1}{N-1} + \frac{8}{N^3} + \frac{12}{N^2}
    - \frac{12}{N} + \frac{8}{(N+1)^3} + \frac{28}{(N+1)^2}
    \right.						\nonumber\\
& & \left. \hspace*{0.6cm}
    -\ \frac{4}{N+1}
    + \frac{32}{3}\frac{1}{(N+2)^2}
    + \frac{224}{9}\frac{1}{N+2}
    \right],						 \\
& &                                                     \nonumber\\
\Delta_{gg}^{(1)}
&=& \frac12 n_f C_F
    \left[
    - \frac{16}{3}\frac{1}{(N-1)^2}
    + \frac{80}{9}\frac{1}{N-1}
    + \frac{8}{N^3}
    - \frac{16}{N^2}
    + \frac{12}{N}
    + \frac{8}{(N+1)^3}
    \right.						\nonumber\\
& & \left. \hspace*{1.3cm}
    -\ \frac{24}{(N+1)^2}
    + \frac{4}{N+1}
    - \frac{16}{3}\frac{1}{(N+2)^2}
    - \frac{224}{9}\frac{1}{N+2}
    \right]						\nonumber\\
&-& \frac43 n_f C_A
    \left[
      S_2(N) - \frac{1}{(N-1)^2} + \frac{1}{N^2}
    - \frac{1}{(N+1)^2} + \frac{1}{(N+2)^2} - \frac{\pi^2}{6}
    \right]						\nonumber\\
&+& C_A^2
    \left[
    - 8 S_1(N) S_2(N)
    + 8 S_1(N)
      \left( \frac{1}{(N-1)^2} - \frac{1}{N^2} + \frac{1}{(N+1)^2}
	   - \frac{1}{(N+2)^2} + \frac{\pi^2}{6}
      \right)
    \right.						\nonumber\\
& & \hspace*{0.7cm}
    + \left( 8 S_2(N) - \frac{4 \pi^2}{3} \right)
      \left( \frac{1}{N-1} - \frac{1}{N} + \frac{1}{N+1}
	   - \frac{1}{N+2} + \frac{11}{12}
      \right)						\nonumber\\
& & \hspace*{0.7cm}
    - \frac{8}{(N-1)^3}
    + \frac{22}{3}\frac{1}{(N-1)^2}
    - \frac{8}{(N-1)^2 N}
    - \frac{8}{(N-1) N^2}
    - \frac{8}{N^3}
    - \frac{14}{3}\frac{1}{N^2}				\nonumber\\
& & \hspace*{0.7cm}
    - \frac{8}{(N+1)^3}
    + \frac{14}{3}\frac{1}{(N+1)^2}
    - \frac{8}{(N+1)^2 (N+2)}
    - \frac{8}{(N+1) (N+2)^2}				\nonumber\\
& & \left. \hspace*{0.7cm}
    - \frac{8}{(N+2)^3}
    - \frac{22}{3}\frac{1}{(N+2)^2}
    \right]						\\
& &                                                     \nonumber
\end{eqnarray}
\end{subequations}
are present specifically for the timelike functions \cite{GRV93}.
In Eqs.~(\ref{e.P1}) the sum
\begin{subequations}
\begin{eqnarray}
S^\prime_{m\pm}(\tfrac{N}{2})
&=& 2^{m-1} \sum_{j=1}^N \frac{1+(-1)^j}{j^m}
\end{eqnarray}
has the analytic continuation
\begin{eqnarray}
S^\prime_{m+}(\tfrac{N}{2}) &\longrightarrow& S_m(\tfrac{N}{2}), \\
S^\prime_{m-}(\tfrac{N}{2}) &\longrightarrow& S_m(\tfrac{N-1}{2}),
\end{eqnarray}
\end{subequations}
with
\begin{eqnarray}
S_3(N) &=& \sum_{j=1}^N \frac{1}{j^3}\
\longrightarrow\ \zeta(3) + \psi^{(2)}_{N+1},		\\
%
%
\widetilde{S}_\pm(N)
&=& -\frac{5}{8} \zeta(3)
\pm \left[
    \frac{S_1(N)}{N^2}
  - \frac{\zeta(2)}{2}
      \big( \psi^{(0)}_{(N+1)/2} - \psi^{(0)}_{N/2} \big)
 +  {\rm Li}(N)
    \right],						\\
\label{e.Stilde}
G^{(1)}(N)
&=& \psi^{(1)}_{(N+1)/2} - \psi^{(1)}_{N/2}.
\end{eqnarray}
The last term in Eq.~(\ref{e.Stilde}) involves an integral over
the dilogarithm function,
\begin{subequations}
\begin{eqnarray}
{\rm Li}(N)
&\equiv& \int_0^1 dx\,x^{N-1}\frac{{\rm Li}_2(x)}{1+x},
\end{eqnarray}
and can be approximated using the expansion \cite{GRV90}
\begin{eqnarray}
{\rm Li}(N)
&\approx& \frac{1.01}{N+1} - \frac{0.846}{N+2} + \frac{1.155}{N+3}
 - \frac{1.074}{N+4} + \frac{0.55}{N+5}\, .
\end{eqnarray}
\end{subequations}

\newpage


\begin{figure}[t]
\includegraphics[width=0.7\textwidth]{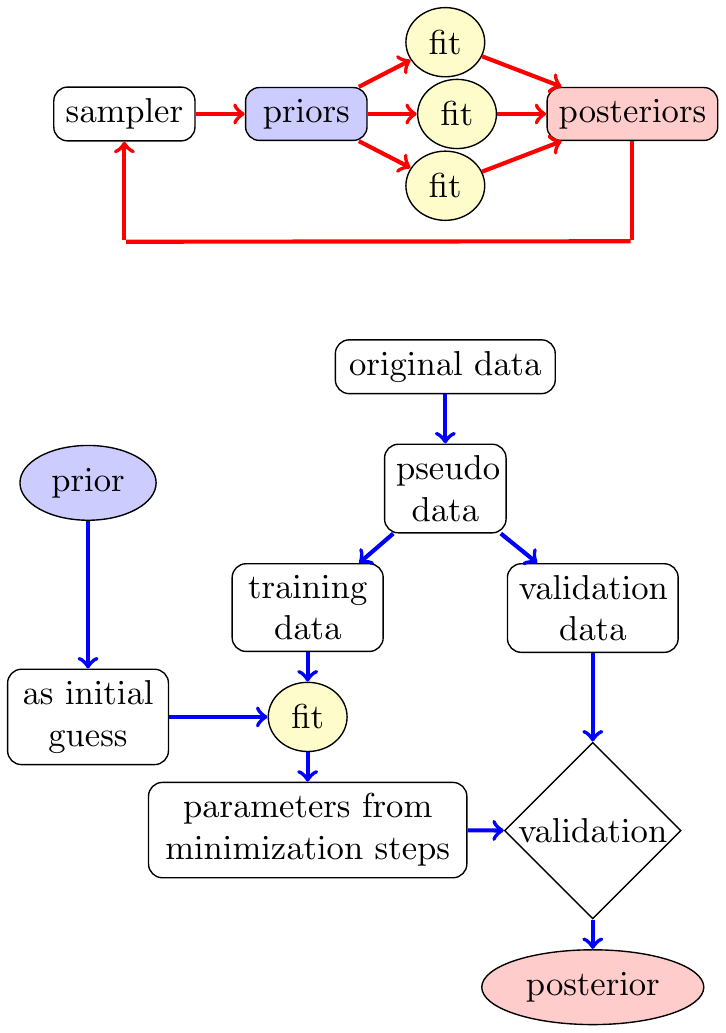}
\caption{Workflow of the iterative Monte Carlo fitting strategy.
	In the upper diagram (red lines) an iteration begins at
	the prior sampler and a given number of fits are performed
	generating an ensemble of posteriors.  After the initial
	iteration, with a flat sampler, the generated posteriors
	are used to construct a multivariate Gaussian sampler for
	the next iteration.
	The lower diagram (with blue lines) summarizes the workflow
	that transforms a given prior into a final posterior.}
\label{f.workflow}
\end{figure}

\begin{figure}[t]
\includegraphics[width=0.6\textwidth]{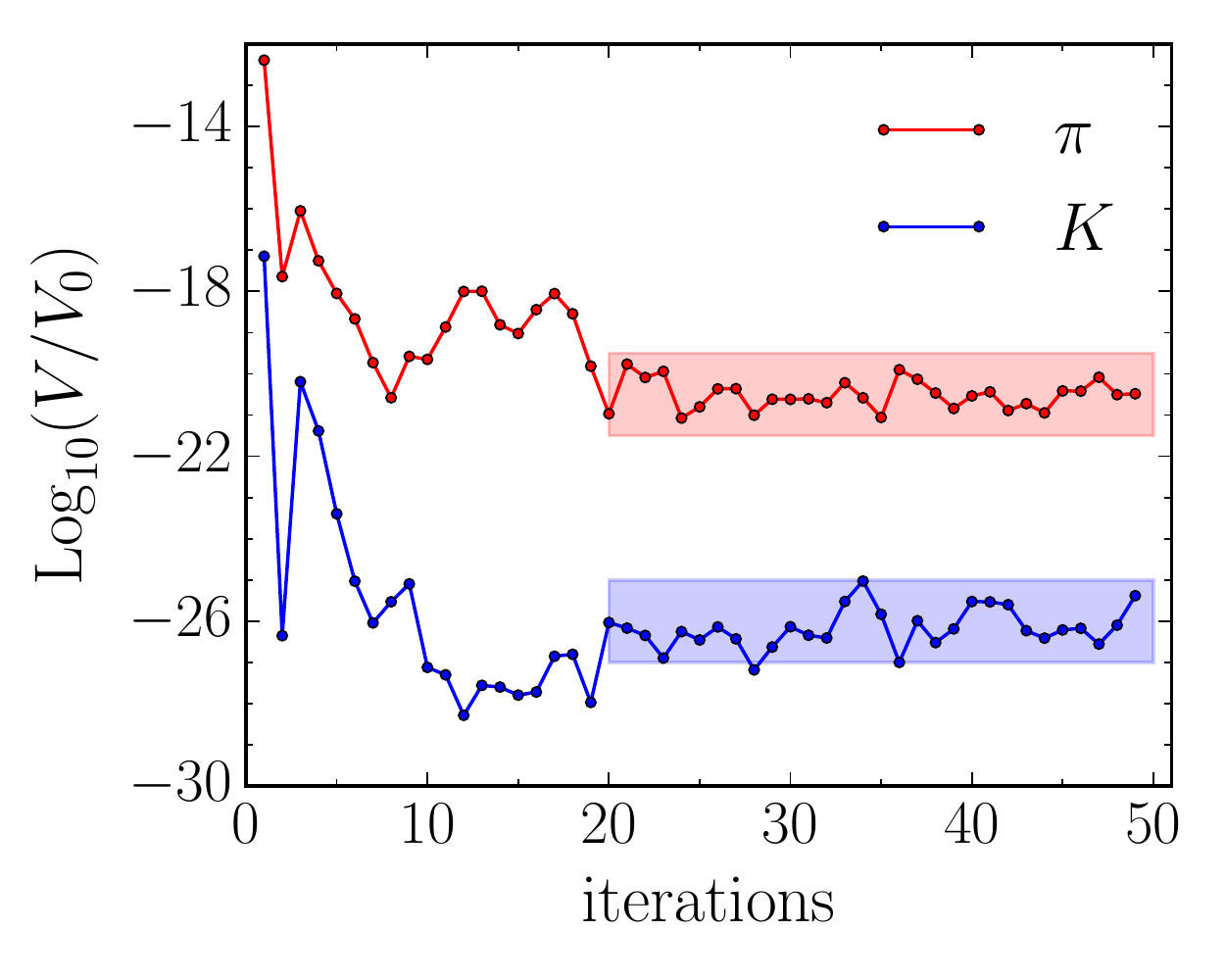}
\caption{Normalized IMC volume versus number of iterations
	for pions (red lines) and kaons (blue lines). The approximate 
  convergence of the volumes are indicated by the colored regions.}
\label{f.vol}
\end{figure}

\begin{figure}[t]
\includegraphics[width=1\textwidth]{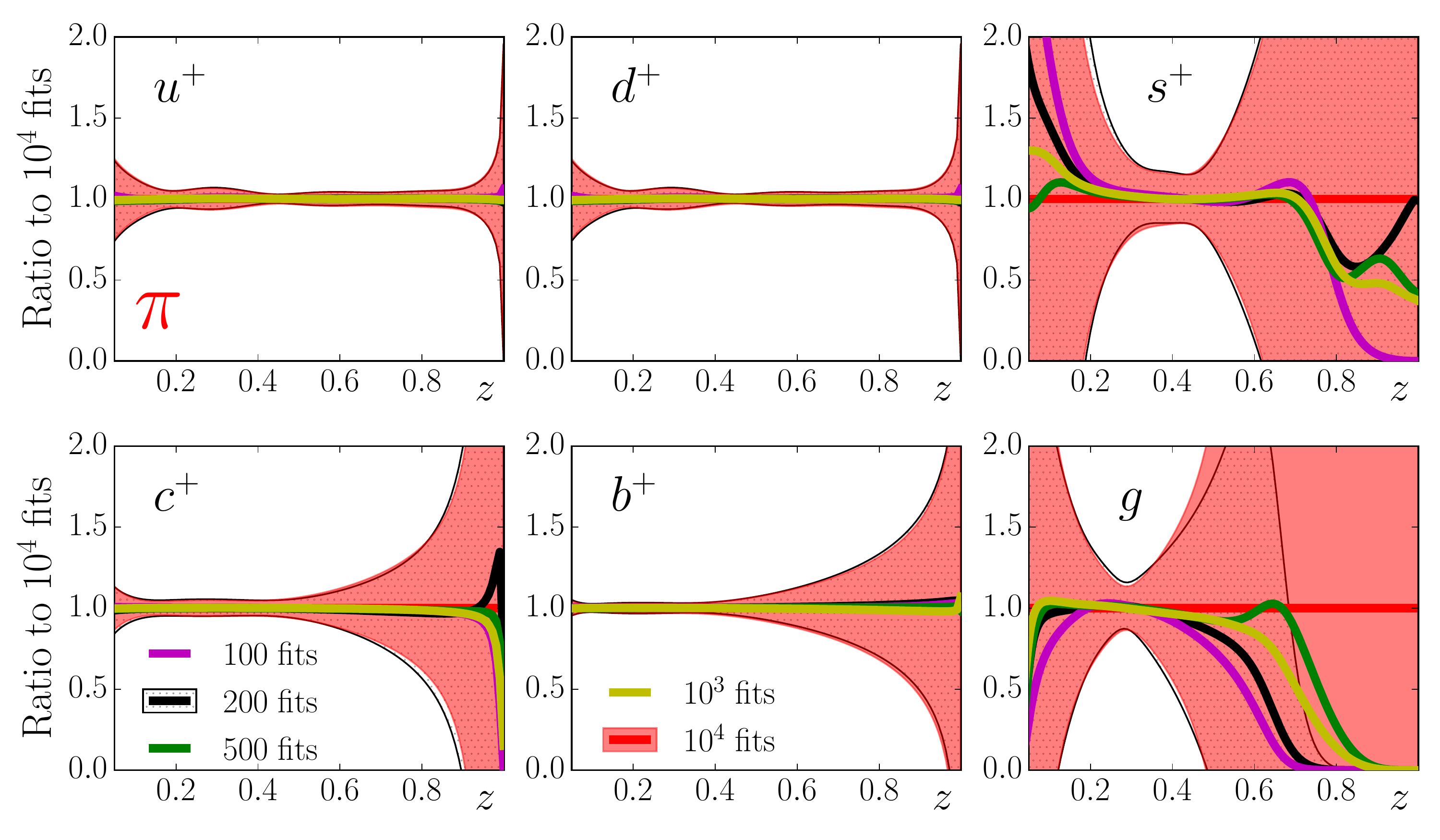}
\includegraphics[width=1\textwidth]{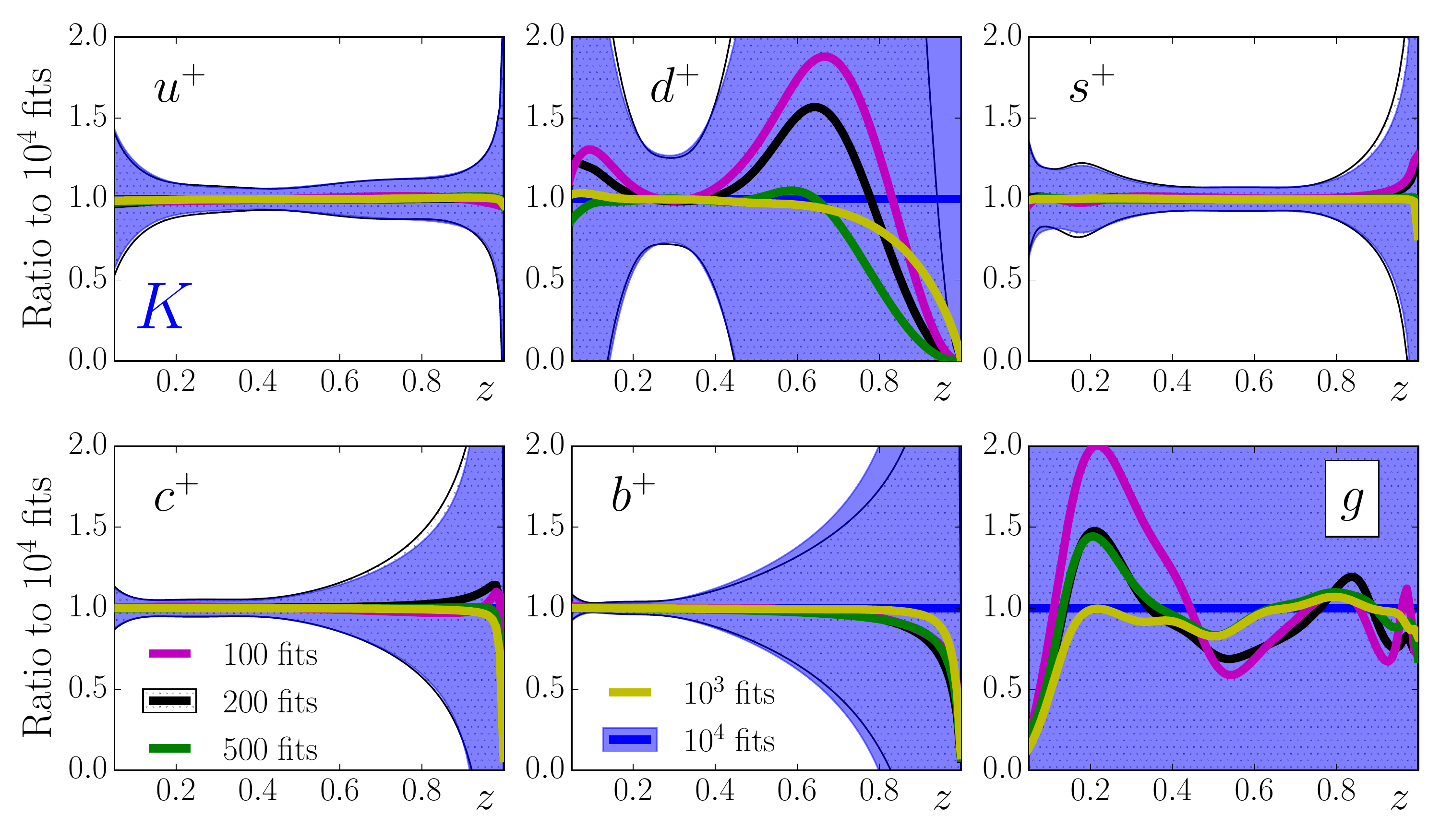}
\caption{Fragmentation functions computed from
	100 (pink), 200 (black), 500 (green), $10^3$ (yellow)
	and $10^4$ (red for pions, blue for kaons) fits,
	normalized to the latter.
	The uncertainties for the 200 (black shaded) and $10^4$
	results are indicated by the bands.}
\label{f.convergence}
\end{figure}

\begin{figure}[t]
\includegraphics[width=1\textwidth]{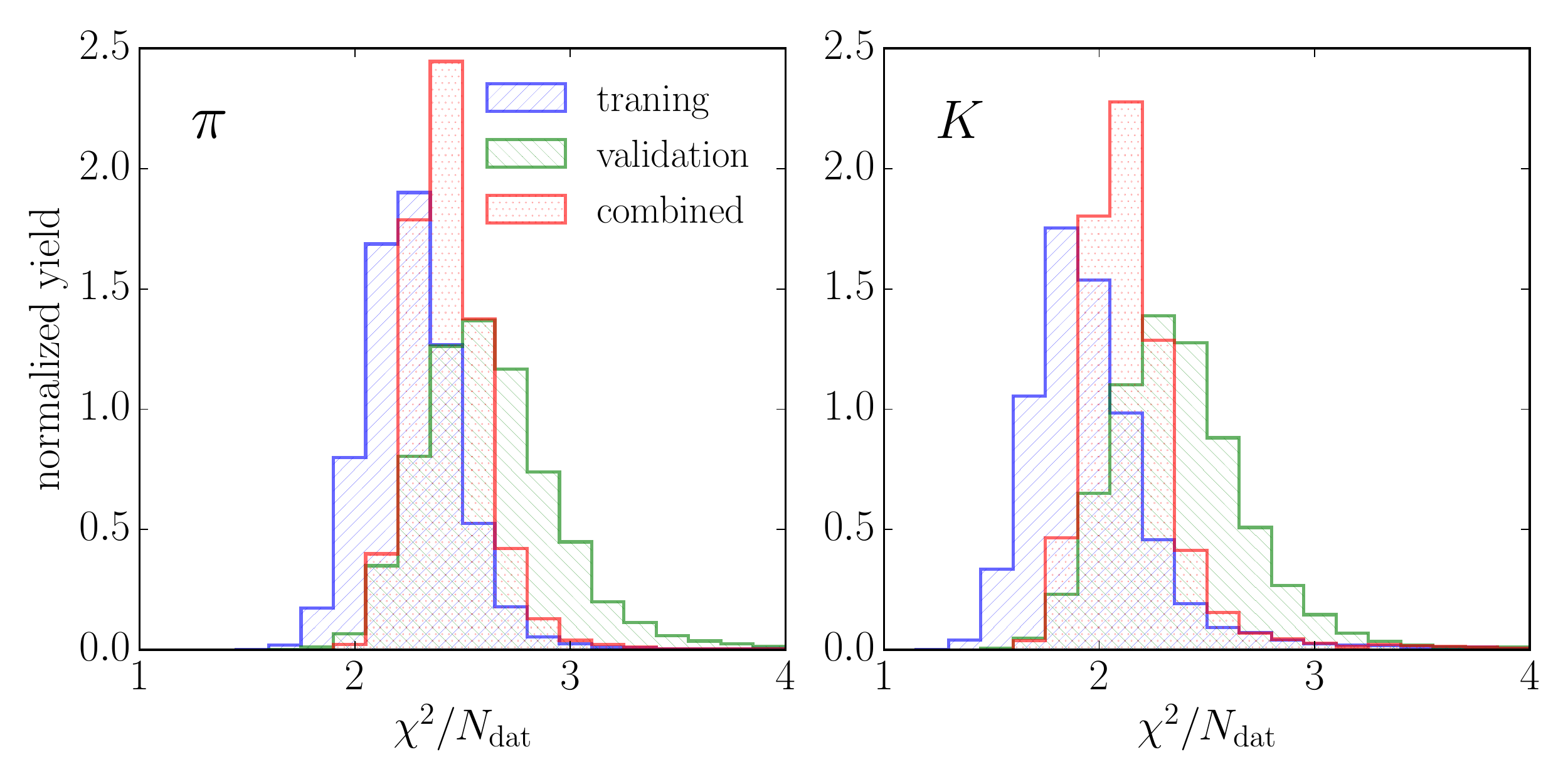}
\caption{Normalized yield of IMC fits versus $\chi^2/N_{\rm dat}$
	for the training (blue forward hashed),
	validation (green backward hashed), and
	combined (red dotted) samples for
	$\pi$ (left panel) and $K$ production (right panel).}
\label{f.chi2}
\end{figure}

\begin{figure}[t]
\includegraphics[width=1\textwidth]{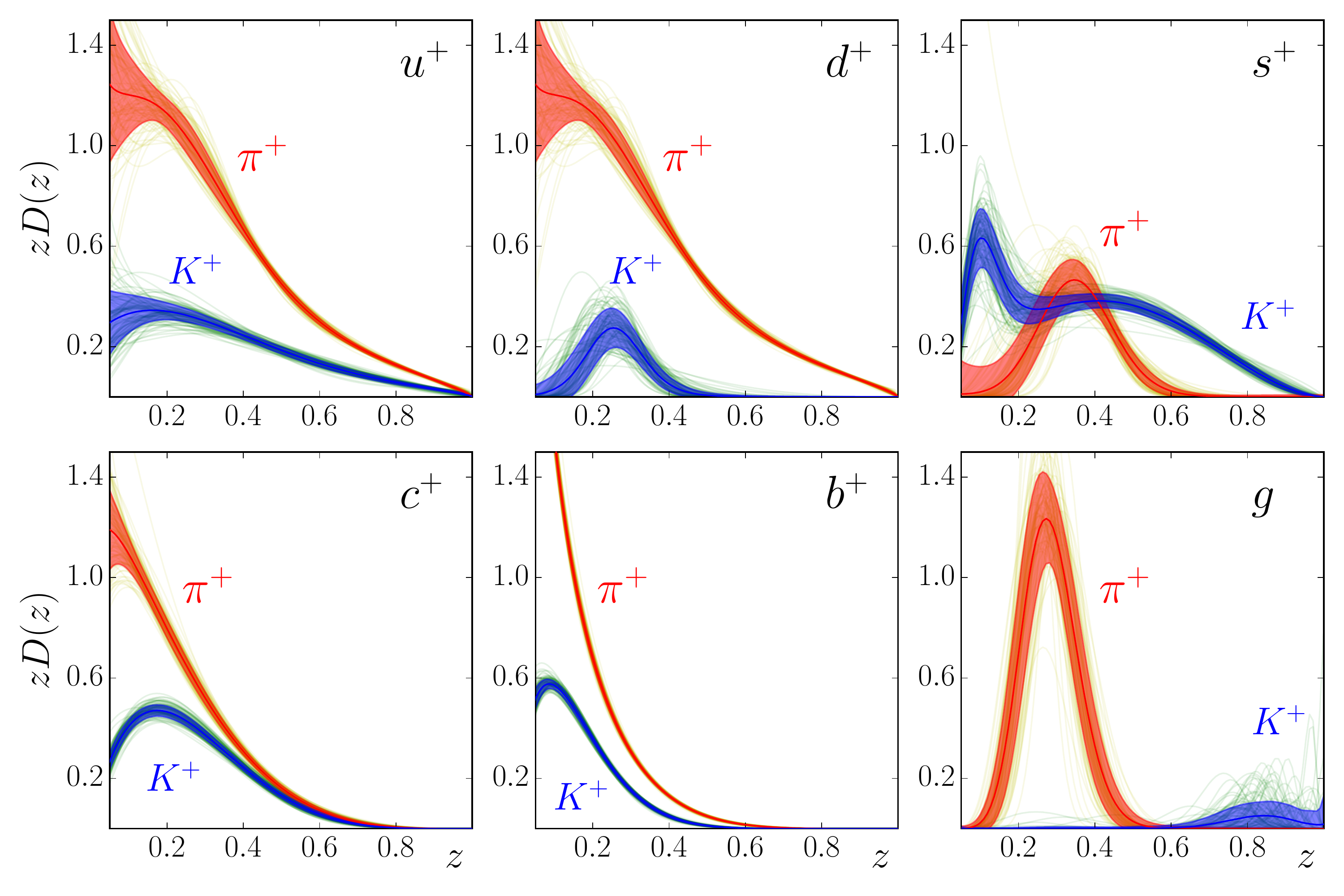}
\caption{Fragmentation functions for $u^+$, $d^+$, $s^+$, $c^+$,
	$b^+$ and $g$ into $\pi^+$ (red bands) and $K^+$ (blue bands)
	mesons as a function of $z$ at the input scale
	($Q^2=1$~GeV$^2$ for light quark flavors and gluon,
	$Q^2=m_q^2$ for the heavy quarks $q=c$ and $b$).
	A random sample of 100 posteriors
	(yellow curves for $\pi^+$, green for $K^+$) is shown
	together with the mean and variance (red and blue bands).}
\label{f.FFQ20}
\end{figure}

\begin{figure}[t]
\includegraphics[width=1\textwidth]{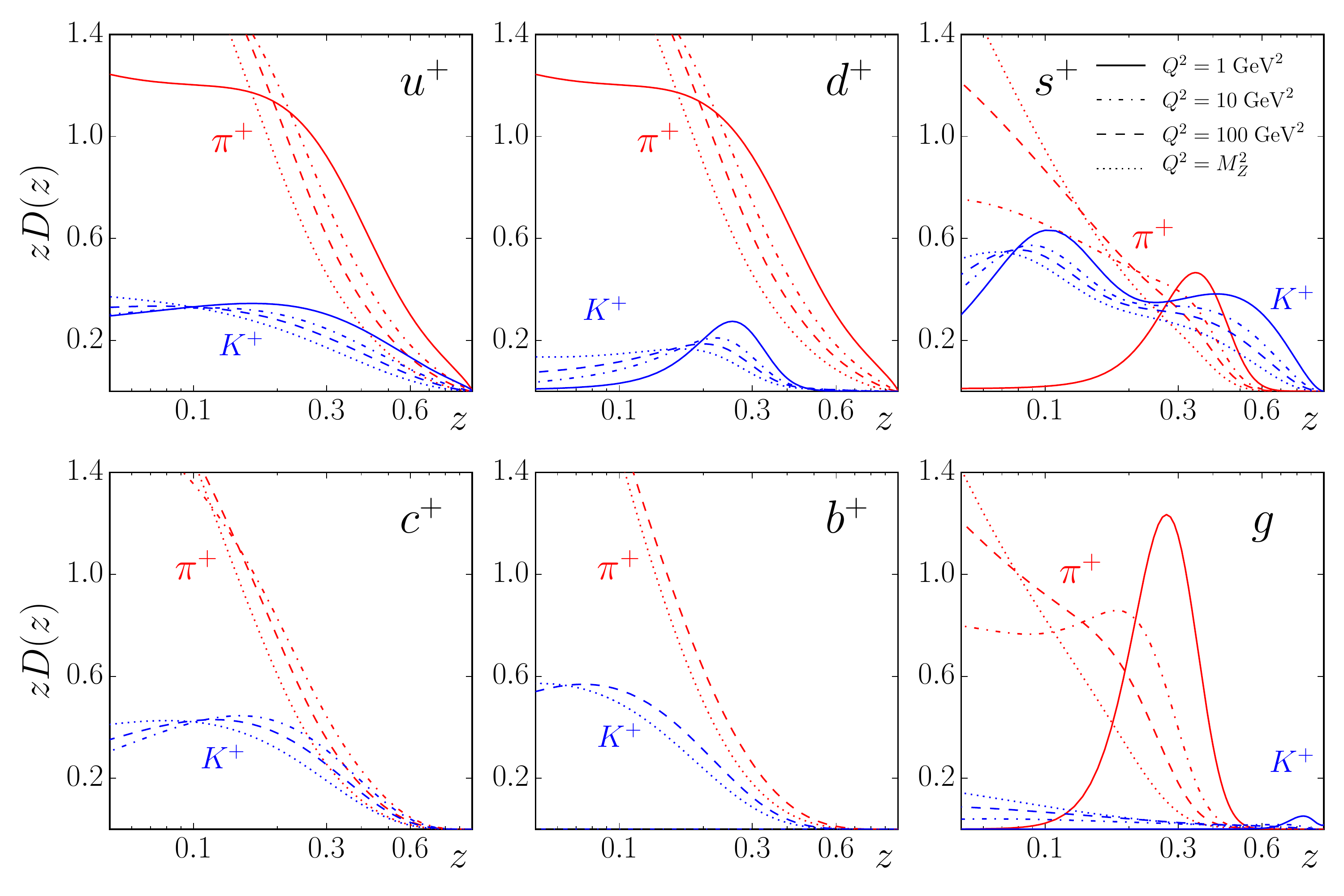}
\caption{Evolution of the $u^+$, $d^+$, $s^+$, $c^+$, $b^+$
	and $g$ fragmentation functions to $\pi^+$ (red curves)
	and $K^+$ (blue curves) with the scale, from the input scale
	$Q^2=1$~GeV$^2$ (solid) to $Q^2=10$~GeV$^2$ (dot-dashed),
	$Q^2=100$~GeV$^2$ (dashed) and $Q^2=M_Z^2$ (dotted).}
\label{f.FFevo}
\end{figure}

\end{document}

%% file: chi2-tab.tex
\begin{tabular}{lccr|ccr|ccr}
experiment  &~ref.~                           &observable &$Q~~~$&\multicolumn{3}{c|}{pions~~~~}                     &\multicolumn{3}{c}{kaons}\\
            &                                 &           & (GeV)& ~~$~N_{\rm dat}$~~ & ~norm.      &$\chi^2$~~~~      & ~$N_{\rm dat}$~~& ~norm.&$\chi^2$\\\hline
ARGUS       & \cite{ARGUS89}                  & inclusive & 9.98 &  35              & 1.024(1.058)& 51.1(55.8)   & 15              & 1.007 & 8.5   \\
Belle       & \cite{Belle13, Leitgab13}       & inclusive & 10.52&  78              & 0.900(0.919)& 37.6(21.7)   & 78              & 0.988 & 10.9  \\
BaBaR       & \cite{BaBar13}                  & inclusive & 10.54&  39              & 0.993(0.948)& 31.6(70.7)   & 30              & 0.992 & 4.9   \\
TASSO       & \cite{TASSO80, TASSO83, TASSO89}& inclusive & 12-44&  29              & $(*)$       & 37.0(38.8)   & 18              & $(*)$ & 14.3  \\
TPC         & \cite{TPC84, TPC86, TPC88}      & inclusive & 29.00&  18              & 1           & 36.3(57.8)   & 16              & 1     & 47.8  \\
            &                                 & $uds$ tag & 29.00&   6              & 1           &  3.7( 4.6)   &                 &       &       \\
            &                                 & $b$ tag   & 29.00&   6              & 1           &  8.7( 8.6)   &                 &       &       \\
            &                                 & $c$ tag   & 29.00&   6              & 1           &  3.3( 3.0)   &                 &       &       \\
HRS         & \cite{HRS87}                    & inclusive & 29.00&   2              & 1           &  4.2( 6.2)   &  3              & 1     & 0.3   \\
TOPAZ       & \cite{TOPAZ95}                  & inclusive & 58.00&   4              & 1           &  4.8( 6.3)   &  3              & 1     & 0.9   \\
OPAL        & \cite{OPAL94, OPAL00}           & inclusive & 91.20&  22              & 1           & 33.3(37.2)   & 10              & 1     & 6.3   \\
            &                                 & $u$ tag   & 91.20&   5              & 1.203(1.203)&  6.6( 8.1)   &  5              & 1.185 & 2.1   \\
            &                                 & $d$ tag   & 91.20&   5              & 1.204(1.203)&  6.1( 7.6)   &  5              & 1.075 & 0.6   \\
            &                                 & $s$ tag   & 91.20&   5              & 1.126(1.200)& 14.4(11.0)   &  5              & 1.173 & 1.5   \\
            &                                 & $c$ tag   & 91.20&   5              & 1.174(1.323)& 10.7( 6.1)   &  5              & 1.169 & 13.2  \\
            &                                 & $b$ tag   & 91.20&   5              & 1.218(1.209)& 34.2(36.6)   &  4              & 1.177 & 10.9  \\
ALEPH       & \cite{ALEPH95}                  & inclusive & 91.20&  22              & 0.987(0.989)& 15.6(20.4)   & 18              & 1.008 & 6.1   \\
DELPHI      & \cite{DELPHI95, DELPHI98}       & inclusive & 91.20&  17              & 1           & 21.0(20.2)   & 27              & 1     & 3.9   \\
            &                                 & $uds$ tag & 91.20&  17              & 1           & 13.3(13.4)   & 17              & 1     & 22.5  \\
            &                                 & $b$   tag & 91.20&  17              & 1           & 41.9(42.9)   & 17              & 1     & 9.1   \\
SLD         & \cite{SLD04}                    & inclusive & 91.28&  29              & 1.002(1.004)& 27.3(36.3)   & 29              & 0.994 & 14.3  \\
            &                                 & $uds$ tag & 91.28&  29              & 1.003(1.004)& 51.7(55.6)   & 29              & 0.994 & 42.6  \\
            &                                 & $c$   tag & 91.28&  29              & 0.998(1.001)& 30.2(40.4)   & 29              & 1.000 & 31.7  \\
            &                                 & $b$   tag & 91.28&  29              & 1.005(1.005)& 74.6(61.9)   & 28              & 0.992 & 134.1 \\\hline
{\bf TOTAL:}&                                 &           &      & 459~             &             & 599.3(671.2)~&391~             &       & 395.0 \\
            &                                 &           &      &\multicolumn{3}{r|}{$\chi^2/N_{\rm dat}=$1.31(1.46)}&\multicolumn{3}{r}{$\chi^2/N_{\rm dat}=$1.01}
\end{tabular}